\documentclass[amsmath,amssymb,pre,showpacs,floatfix,twocolumn]{revtex4}
\usepackage{graphicx}
\usepackage{dcolumn}

\begin{document}

\title{Chiral Brownian rotor and heat pump}
\author{M. van den Broek}
\author{C. Van den Broeck}
\affiliation{Hasselt University, B-3590 Diepenbeek, Belgium}
\begin{abstract}
This paper provides some comprehensive calculations supporting the results in Phys.~Rev.~Lett.~\textbf{100}, 130601 (2008).
\end{abstract}


\maketitle

%%%%%%%%%%%%%%%%%%%%%%%%%%%%%%%%%%%%%%%%%%%%%%
\section{Introduction}
%%%%%%%%%%%%%%%%%%%%%%%%%%%%%%%%%%%%%%%%%%%%%%

Brownian motors have been studied intensively since the early 1990s \cite{Reimann, Astumian, Julicher,Leibler}. This interest coincided with developments in bioengineering and nano\-techno\-logy, where understanding and designing a motor in the shape of a small biological or artificial device is an important issue. Most of the motors investigated in this context are powered by chemical energy. Brownian motors driven by a temperature gradient  \cite{Smoluchowski, Landauer,motor,Meurs1,Meurs2,Meurs3} have a fundamental appeal, since their operation is directly related to basic questions such as Carnot efficiency, Maxwell demons and the foundations of statistical mechanics and thermodynamics \cite{Parrondo,Sekimoto,Jarzynski,ACP,Astumian2}. 
The additional significance of the thermal Brownian motor comes from the recent observation that it can operate as a refrigerator \cite{refrigerator,Nakagawa}, see also \cite{pekola}. In fact, this property is, at least in the regime of linear response, a direct consequence of Onsager symmetry \cite{onsager}: if a temperature gradient generates motion, an applied force will generate a heat flux. This principle is well known in its application to electro-thermal devices, displaying  the Peltier, Seebeck and Thompson effects \cite{callen}. At variance however with these macroscopic devices, rectification of nonequilibrium thermal fluctuations provide the driving mechanism for Brownian refrigeration. The latter become more prominent, and so do the resulting motor and cooling functions, as the apparatus becomes smaller. 

Since the properties of the Brownian heat pump follow by Onsager symmetry from those of the Brownian motor, we first focus on the latter. 

%%%%%%%%%%%%%%%%%%%%%%%%%%%%%%%%%%%%%%%%%%%%%%
\section{Brownian motors}
%%%%%%%%%%%%%%%%%%%%%%%%%%%%%%%%%%%%%%%%%%%%%%

In earlier suggestions, Brownian motors move linearly, which obviously poses difficulties when comparing with real systems, or suggesting a technological implementation of a Brownian motor. In this paper we introduce a Brownian motor, driven by thermal fluctuations, that is free to rotate around a fixed axis. Rotational motion typically encounters less friction than purely translational movement and a rotating force is easier to apply than a linear force. We propose a device that exploits the random nature of the perturbations from its environment maximally to produce a net directed motion.

Molecular motors operating within biological cells, although chemically driven, are also subject to random motion.
\begin{figure}
    \begin{center}
        \includegraphics[width=0.6\columnwidth]{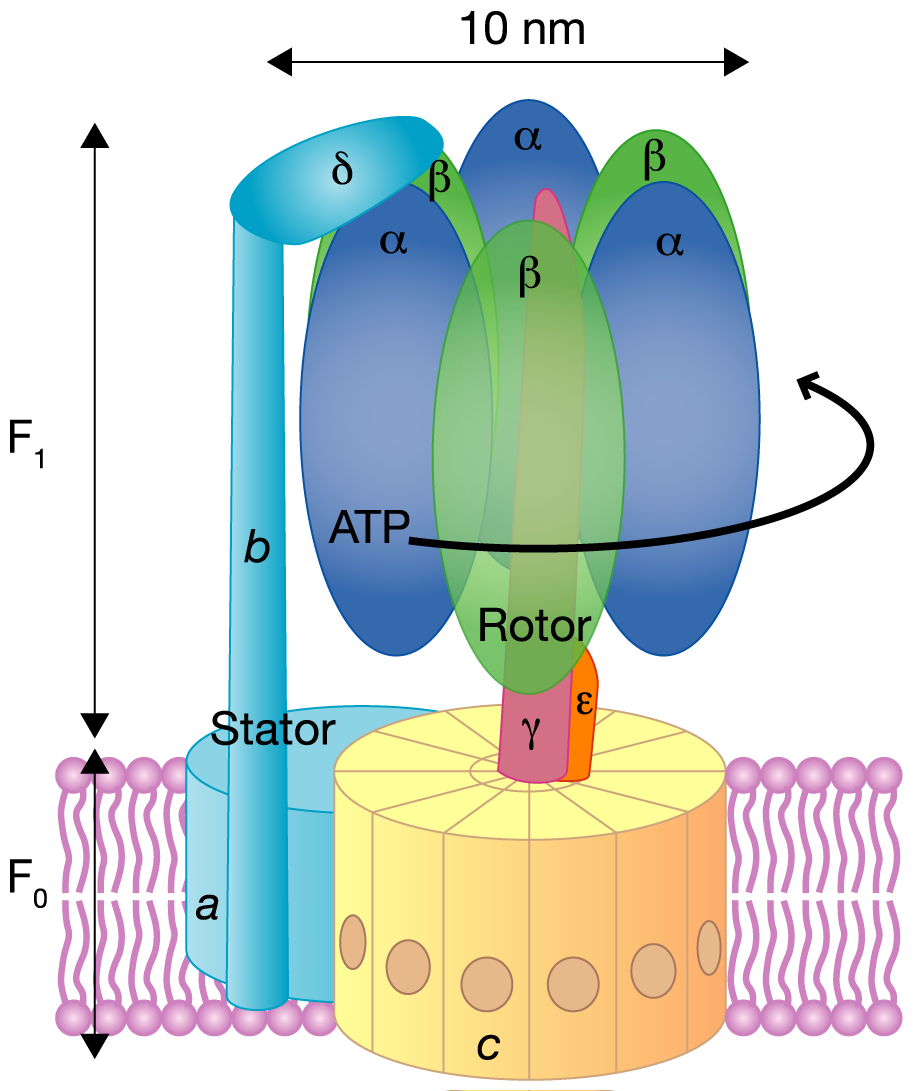}
    \end{center}
    \caption{F$_1$-ATPase is a naturally occurring rotating molecular motor. It works as a pump for ions through a membrane. Its driving force is chemical, through the hydrolysis of ATP. Our interest in this molecular motor lies in its environment (in the cytoplasm, subject to fluctuations), its construction (rotational, and through a biological membrane), its physical characteristics (size of the order of 10 nm) and dynamical properties (rotational frequency of the order of 100 Hz).
\label{fig:F1ATPase}}
\end{figure}
F$_1$-ATPase is a well-known rotating motor (see Fig.\:\ref{fig:F1ATPase} for a diagram). A direct observation of its rotation, driven by the hydrolysis of adenosine triphosphate (ATP), was first reported in \cite{Noji,Yasuda}. Later experiments \cite{Itoh} revealed the direction in which the ATP motor spins.
It is about 10\,nm in size and typically rotates with a frequency of 100\,Hz.  
The observed rotary torque reaches more than 40\,pN\,nm.
The relation between the geometry of the rotating object, specifically its chirality, and its kinetic properties, such as the average motion and friction can be of interest to microbiology. One might also imagine artificial devices inspired by the existing biological examples. Proteins could be used as the building blocks of mechanical devices and artificial biological membranes as means to separate reservoirs and keep them at different temperature. Small moving parts in the area of micro-electronics are also subject to random fluctuations.

\begin{figure*}
    \begin{center}
        \includegraphics[width=0.6\textwidth]{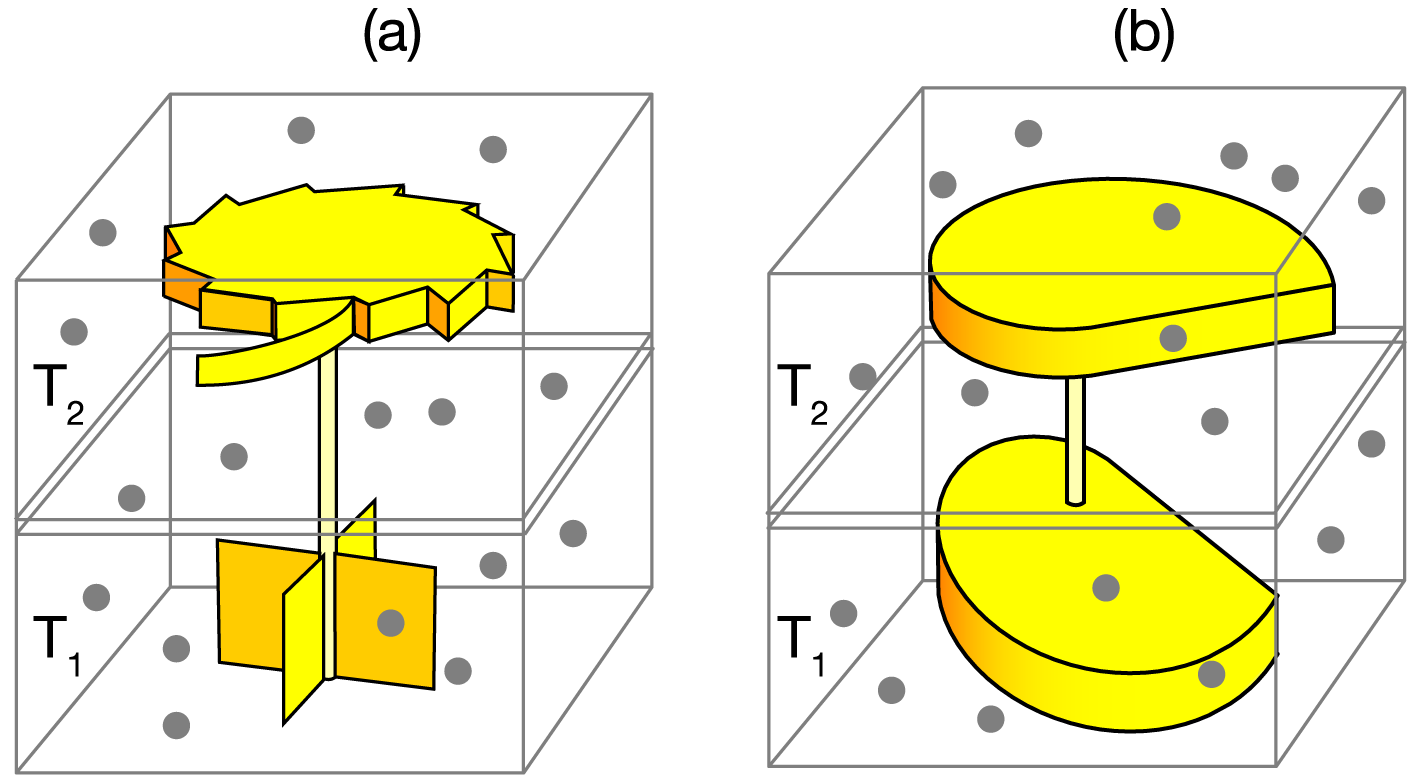}
    \end{center}
    \caption{(a) The ratchet and pawl mechanism used by R.\:Feynman to illustrate the impossibility to extract work from a system in equilibrium. An axle with vanes in it (in the right reservoir ) is bombarded by gas molecules at temperature $T_{1}$.   The pawl in the left reservoir (surrounded by a gas at temperature $T_{2}$) \emph{seems} to allow only one rotation sense of the ratchet, that is connected with the axle. The fluctuations of the paddle in the right reservoir \emph{would} then be rectified. Via a comparison of the probability to move forward and backward,  Feynman showed in his Lectures on Physics \cite{feynman} that at temperature equilibrium, $T_{1} = T_{2}$, no average motion occurs and the device cannot be used to do work, such as to lift a weight. For $T_{1} > T_{2}$ however, average motion does take place and the ratchet works as an engine. Feynman noted that for $T_{1} < T_{2}$, the ratchet goes backward.  Note that the rectification manifested by the device originates from the asymmetry of the ratchet and pawl mechanism. Our model remains close to Feynman's system, as it can be applied to rotating three-dimensional objects of any shape, while it simplifies the asymmetry requirements to the geometrical properties of the device.
(b) A rotating Brownian motor where two parts in isolated thermal reservoirs (temperatures $T_{1}$ and $T_{2}$, particle densities  $\rho_{1}$ and $\rho_{2}$) are connected through the axis of rotation. Collisions with particles in the reservoirs will cause fluctuating rotational movement of the heavier motor, which under appropriate conditions, will propel the motor with a nonzero average angular velocity. 
\label{fig:motors}}
\end{figure*}

As a side note we remark that the rotational three-dimensional model we will present, can be used to describe the essence of the device R.\:Feynman presented in his Lectures on Physics \cite{feynman} [for a sketch of the ratchet and pawl mechanism, see Fig. \ref{fig:motors}(a)] to illustrate the impossibility of a Maxwell Demon, that would be able to extract work from a system in equilibrium. It was also shown that the same device can lift a weight (do work) at temperature disequilibrium. 

These reasons motivate us to study a chiral Brownian motor in detail. The  constituting parts of the motor are in different thermal reservoirs. In an analytical analysis we will derive expressions for the kinetic properties of the motor as a function of the external parameters of the system. It will be made clear that temperature equilibrium between the different reservoirs results in zero average motion and hence prohibits the creation of a Maxwell Demon. On the other hand we will demonstrate the importance of the configuration of the building blocks of the motor and their actual shape, as well as their position relative to the rotation axis. Some emphasis is put on finding optimum operation, yielding maximum average angular velocity. The chiral Brownian motor presented in this paper is a precursor for the chiral Brownian refrigerator presented in the next. The exact relationship derived here between the angular velocity and the temperature gradient will be a crucial step in the investigation of the cooling potential. 

The concrete model we propose consists of at least two parts, each residing in a thermal reservoir $i = 1, 2, \ldots$, that are rigidly connected with each other through a rotation axis. Fig.\:\ref{fig:motors}(b) shows the construction of two parts in reservoirs of temperature $T_{1}$ and $T_{2}$, and particle densities $\rho_{1}$ and $\rho_{2}$. We expect that the fluctuations from collisions with particles in the thermal reservoirs will under certain conditions be rectified, resulting is an average rotational motion, clockwise or counterclockwise. As we will show, these conditions are (1) thermal disequilibrium, $T_{1} \neq T_{2}$, and (2) asymmetry (or chirality) through the geometrical shape of the motor parts.

%%%%%%%%%%%%%%%%%%%%%%%%%%%%%%%%%%%%%%%%%%%%%%
\section{From fluctuations to the angular velocity} 
%%%%%%%%%%%%%%%%%%%%%%%%%%%%%%%%%%%%%%%%%%%%%%
\label{fluctutions}

Our analysis is based on an exact calculation of the probability for the motor to change its rotating speed by a certain amount when subject to thermal fluctuations. We will show that an exact solution can be reached when the fluctuations are in the form of collisions of particles of an ideal gas at temperature equilibrium with the surface of the motor. A master equation for the probability density to observe an angular velocity $P(\omega, t)$ at a certain time $t$ can be proposed if  the particles are presumed to collide not more than once, and only with the motor. This condition implies that the gas is in the high Knudsen number regime and that the shape of the motor is such that recollisions are impossible. We therefore limit the
parts of the motor to convex and closed shapes.

We are interested in the case where the motor, with total mass $M$, has
no translational degree of freedom and a single rotational degree of freedom. Parts of the motor reside in different thermal reservoirs but are considered rigidly linked.
Choosing the $z$-axis as the axis of rotation, we can write for the angular velocity $\vec{\omega} = (0,
0, \omega)$. The inertial moment $I_{z}$ of the motor with respect to the rotation axis is simply denoted as $I$.

Under these conditions the probability density $P(\omega, t)$ obeys a
master equation,
\begin{equation}
\partial_{t} P(\omega, t) = \int d\omega' \left[W(\omega|\omega') P(\omega', t) - W(\omega'|\omega) P(\omega, t)
\right],
\end{equation}
where $W(\omega|\omega')$ is the transition probability per unit time for the motor to
change its angular velocity from $\omega'$ to $\omega$. 
The solution is based on the van Kampen $1 / \Omega$-method \cite{vankampen}.
A Taylor expansion of the first term of the
integrand in the angular velocity change, $\upsilon  = \omega - \omega'$, leads to
\begin{equation}
\partial_{t} P(\omega, t) = \sum_{n = 1}^{\infty} \frac{(-1)^{n}}{n!}
\left (\frac{d}{d\omega}\right )^{n}\{a_{n}(\omega) P(\omega, t) \}.
\label{evolutionprobability}
\end{equation}
In this expression the so-called \emph{jump moments} appear, given by
\begin{equation}
a_{n}(\omega) = \int \upsilon ^{n} W(\omega;\upsilon) d\upsilon.
\end{equation}
A notation
$W(\omega'; \upsilon) = W(\omega|\omega')$ is used.
With the time evolution of the probability density known (Eq.\:\ref{evolutionprobability}), it is possible to derive a coupled set of equations for the moments of the angular velocity $\langle \omega^{n} \rangle$:
\begin{equation}
\partial_{t} \langle \omega^{n} \rangle = \sum_{k=1}^{n} \binom {n}{k} \langle \omega ^{n -
k} a_{k}(\omega) \rangle,
\label{coupledset}
\end{equation}
with  $\binom {n}{k}$ the binomial coefficients.
Our strategy is now clear: first find an expression for the transition probability $W(\omega|\omega')$, then calculate the jump moments $a_{n}(\omega)$, and finally the moments of the angular velocity $\langle \omega^{n} \rangle$.

However, the coupled set of equations Eq.\:\ref{coupledset} cannot be solved unless we expand each equation into powers of a small variable, and ignore terms after a certain order. For the expansion variable we will use $\varepsilon = r_{0} \sqrt{m/I}$, with $r_{0} = \sqrt{I /M}$ the radius of gyration of the motor. We also introduce an effective temperature $T_{\text{eff}}$, so that to first significant order, in the regime of stationary motion, the average kinetic energy of the motor is given by
\begin{equation}
\frac{1}{2} I \langle \omega^{2} \rangle = \frac{1}{2} k_{B} T_{\text{eff}}.
\end{equation}

In the calculation it is convenient to do a transformation to dimensionless variables, by scaling the angular velocity $\omega$ and the jump moments $a_{n}$ as follows:
\begin{align}
\xi  &= \omega \sqrt{I /
k_{B} T_{\text{eff}}},\nonumber
\\
A_{n}(\xi) &= (\sqrt{I / k_{B} T_{\text{eff}}})^{n} a_{n}(\xi).
\end{align}
Our selfconsistent definition of the effective temperature $T_{\text{eff}}$ then leads to
$\langle \xi^{2} \rangle = 1$ for the stationary state to first order in $\varepsilon$.
The set of coupled equations for the moments $\langle \xi ^{n} \rangle =
\int \xi ^{n} P(\xi, t) d \xi $ remains
\begin{equation}
\partial_{t} \langle \xi ^{n} \rangle = \sum_{k=1}^{n} \binom {n}{k} \langle \xi ^{n -
k} A_{k}(\xi) \rangle.
\end{equation}

%%%%%%%%%%%%%%%%%%%%%%%%%%%%%%%%%%%%%%%%%%%%%%
\section{Two-dimensional model of the motor}
%%%%%%%%%%%%%%%%%%%%%%%%%%%%%%%%%%%%%%%%%%%%%%

\begin{figure}
    \begin{center}
        \includegraphics[width=0.7\columnwidth]{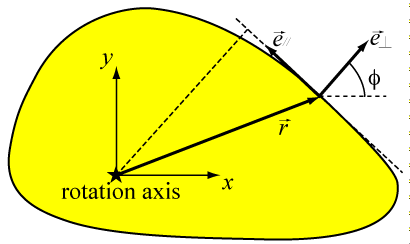}
    \end{center}
    \caption{In each reservoir, the motor part can be described by the boundary $\vec{r}(x,y)$, where the rotation axis is at the origin of the reference frame. It is convenient to also know explicitly the orientation of the boundary at any point. This is given by the polar angle $\varphi$ of the normal outward vector $\vec{e}_{\perp}$ on the surface at this point. The unit vector $\vec{e}_{\shortparallel}$ is tangential to the boundary.
    \label{fig:coordinates}}
\end{figure}

The motor consists of parts with hard surfaces of arbitrary (but convex) shape, each described by their boundary and inner mass distribution.
Many of the important features already appear in a simpler two-dimensional system, which we present first.
Here the motor consists of two-dimensional shapes, each in two-dimensional reservoirs.
We choose a cartesian coordinate system as follows: the $z$-axis coincides with the rotation axis, while the $xy$-plane is parallel to the reservoirs.
In each reservoir $i$, the shape of the motor (part) is defined by its boundary $\vec{r}_{i}(x,y)$, given as a vector with the rotation axis as its origin (see Fig.\:\ref{fig:coordinates}).
 The perimeter of the boundary is denoted $L_{i}$. Henceforth we will just write $\vec{r}(x,y)$ for $\vec{r}_{i}(x,y)$ as no confusion can arise in subsequent expressions.
It is convenient in the further derivation to explicitly know the inclination of the boundary at $\vec{r}(x,y)$, for which we use the orientation of the normal
outward unit vector $\vec{e}_{\perp}  = (\cos
\varphi, \sin \varphi)$ on the boundary, determined by the polar coordinate $\varphi$ from the  $x$-axis.

%%%%%%%%%%%%%%%%%%%%%%%%%%%%%%%%%%%%%%%%%%%%%%
\subsection{Conservation rules}
%%%%%%%%%%%%%%%%%%%%%%%%%%%%%%%%%%%%%%%%%%%%%%
Fluctuations of the angular velocity $\omega$ of motor arise from collisions on the surface with gas particles of mass $m$. Such a collision -- presumed instantaneous
and perfectly elastic -- changes the velocity of the gas particle  $\vec{v'} = (v'_{x}, v'_{y})$  into $\vec{v} = (v_{x}, v_{y})$ after the collision, while the motor changes angular speed from $\omega'$ to $\omega$. The inertial moment of the motor about the rotation axis ($z$-axis) is denoted $I$, while its mass is $M$. 
Conservation of the total energy requires that
\begin{subequations}
\begin{equation}
\frac{1}{2}I{\omega'}^2 + \frac{1}{2}m {v'_{x}}^2 + \frac{1}{2}m {v'_{y}}^2  = \frac{1}{2}I \omega ^2 +
\frac{1}{2}m v_{x}^2 + \frac{1}{2}m v_{y}^2 ,
\end{equation}
while conservation of the angular momentum in the $z$-direction yields
\begin{equation}
m (x v'_{y} - y v'_{x}) + I \omega' = m (x v_{y} - y v_{x}) + I \omega.
\end{equation}
Also we suppose the interaction force is short-ranged and central, implying that the
tangent component of the momentum of the gas particle on the boundary is conserved. Choosing the tangent unit vector $\vec{e}_{\shortparallel} = (-\sin \varphi, \cos \varphi)$, so that $(\vec{e}_{\perp}, \vec{e}_{\shortparallel})$ forms a positive orthonormal base, we write
\begin{equation}
\vec{v'} \cdot \vec{e}_{\shortparallel} = \vec{v} \cdot \vec{e}_{\shortparallel}.
\end{equation}
\label{eq:conservationlaws}
\end{subequations}
The conservation laws (Eq.÷\ref{eq:conservationlaws}) produce a solution for the postcollisional angular velocity $\omega$,
\begin{equation}\label{eq:postcollisional}
\omega
= \omega' +  \frac{2 (\omega' y + v'_{x}) \cos \varphi -  2 (\omega' x - v'_{y}) \sin \varphi}
{ x \sin \varphi - y \cos \varphi + \frac{I}{m} \left(x \sin \varphi - y \cos \varphi \right)^{-1}},
\end{equation}
Introducing
\begin{equation}
r_{\shortparallel} = \vec{r} \cdot \vec{e}_{\shortparallel} = - x \sin \varphi + y \cos \varphi,
\end{equation}
and the precollisional speed of the boundary at position $\vec{r}$,
\begin{equation}
\vec{V'} = \vec{\omega'} \times \vec{r} = (- \omega' y, \omega' x),
\end{equation}
so that we can write
\begin{equation}
V'_{\perp} = \vec{V'} \cdot \vec{e}_{\perp} = - \omega' y  \cos \varphi +  \omega' x \sin \varphi,
\end{equation}
and
\begin{equation}
\Delta V'_{\perp} = (\vec{V'} - \vec{v'}) \cdot \vec{e}_{\perp} = -(\omega' y + v'_{x}) \cos \varphi +  (\omega' x - v'_{y}) \sin \varphi,
\end{equation}
the transition in $\omega$ can also be written as
\begin{equation}
\omega = \omega' + 2 \frac{\Delta V'_{\perp}}{r_{\shortparallel} + \frac{I}{m r_{\shortparallel}}}  .
\end{equation}

%%%%%%%%%%%%%%%%%%%%%%%%%%%%%%%%%%%%%%%%%%%%%%
\subsection{Transition probability}
%%%%%%%%%%%%%%%%%%%%%%%%%%%%%%%%%%%%%%%%%%%%%%
Next, we set out to find the crucial transition probability $W(\omega|\omega')$ for the motor to
change its angular velocity from $\omega'$ to $\omega$ in a unit of time.
Every reservoir $i$ contains a gas with
particle density $\rho_{i}$ and  velocity distribution
$\phi_{i}$. The contribution $dW_{i}$ to the total transition probability $W(\omega | \omega')$ from all possible collisions of particles in gas $i$ in a time interval $dt$ on a
boundary section of length $dl_{i}$, at position $\vec{r}(x,y)$ and with orientation $\varphi$, can be expressed as
\begin{multline}
dW_{i}(\omega | \omega') = dl_{i} \int_{-\infty}^{+\infty} \negthickspace \negthickspace dv'_{x} \int_{-\infty}^{+\infty} \negthickspace \negthickspace dv'_{y}
\\
\times H[(\vec{V'} - \vec{v'}) \cdot \vec{e}_{\perp}]
 |(\vec{V'} - \vec{v'})
\cdot
\vec{e}_{\perp}| 
 \rho_{i} \phi_{i}(v'_{x}, v'_{y})
 \\
\times \delta \left[\omega - \omega' -  \frac{2 (\omega' y + v'_{x}) \cos \varphi -  2 (\omega' x - v'_{y}) \sin \varphi}
{ x \sin \varphi - y \cos \varphi + \frac{I}{m} \left(x \sin \varphi - y \cos \varphi \right)^{-1}}
\right ],
\end{multline}
with $H$ the Heaviside step function and $\delta$ Dirac's distribution.
We multiplied the particle density $\rho_{i}$ with the volume of the gas that is passed by the boundary element  $dl_{i}$ in a time unit, considering only those gas particles that comply with the collision rules.
This can be written in short form as
\begin{multline}
dW_{i}(\omega | \omega') = dl_{i} \int_{-\infty}^{+\infty} \negthickspace \negthickspace dv'_{x} \int_{-\infty}^{+\infty} \negthickspace \negthickspace dv'_{y}
 H[\Delta V'_{\perp}]
 |\Delta V'_{\perp}|
 \\
\times \rho_{i} \phi_{i}(v'_{x}, v'_{y}) \delta \left[\omega - \omega' -  2 \frac{\Delta V'_{\perp}}{r_{\shortparallel} + \frac{I}{m r_{\shortparallel}}}
\right].
\end{multline}

The \emph{total} transition probability is then found by integrating over all boundary elements $dl_{i}$ and summing over all reservoirs:
\begin{equation}
W(\omega | \omega') = \sum_{i} \oint_{\text{boundary}}\negthickspace \negthickspace \negthickspace \negthickspace \negthickspace  \negthickspace \negthickspace \negthickspace  \negthickspace \negthickspace dW_{i}(\omega | \omega').
\end{equation}
Henceforth we will simply write $\oint$ when we imply the line integral over all boundary elements.

For a Maxwellian velocity distribution
at temperature $T_{i}$,
\begin{equation}
\phi_{i}(v_{x}, v_{y}) = \frac{m}{2 \pi k_{B} T_{i}} \exp \left(\frac{- m (v_{x}^2 + v_{y}^2}{2 k_{B} T_{i}}\right
), \label{eq:maxwellian}
\end{equation}
the integrals over the speed of the colliding particles can be performed
explicitly, resulting in
\begin{multline} W(\omega | \omega') = \frac{1}{4} \sum_{i}
\oint dl_{i} \rho_{i}
\sqrt{\frac{m}{2 \pi k_{B} T_{i}}}
\\
\times ( r_{\shortparallel} + \frac{I}{m r_{\shortparallel}})^{2}
H[(\omega - \omega') r_{\shortparallel}] |\omega -
\omega'|
\\
\times \exp \left[- \frac{m}{2 k_{B} T_{i}}
 \left((r_{\shortparallel} + \frac{I}{m r_{\shortparallel}}) \frac{\omega' - \omega}{2} - r_{\shortparallel} \omega' \right)^{2}
\right ]. \label{transitionprobability}
\end{multline}

%%%%%%%%%%%%%%%%%%%%%%%%%%%%%%%%%%%%%%%%%%%%%%
\subsection{Moments of the angular velocity}
%%%%%%%%%%%%%%%%%%%%%%%%%%%%%%%%%%%%%%%%%%%%%%
Now that we have obtained an exact expression for the transition probability $W(\omega|\omega')$, we turn our attention to the jump moments,
\begin{equation}
a_{n}(\omega) = \int \upsilon ^{n} W(\omega;\upsilon) d\upsilon,
\end{equation}
and then the moments of the angular velocity.
Careful consideration of the sign of $r_{\shortparallel} + I / m r_{\shortparallel}$ in
\begin{multline}
W(\omega; \upsilon) = 
\frac{1}{4} \sum_{i}
 \rho_{i}
\sqrt{\frac{m}{2 \pi k_{B} T_{i}}} 
\\
\times \left(H[\upsilon] \int_{r_{\shortparallel}\ge 0} dl_{i} + H[-\upsilon] \int_{r_{\shortparallel}<0} dl_{i}\right)
 | \upsilon |
( r_{\shortparallel} + \frac{I}{m r_{\shortparallel}})^{2}
\\
\times \exp \left[- \frac{m}{2 k_{B} T_{i}}
\left((r_{\shortparallel} + \frac{I}{m r_{\shortparallel}}) \frac{\upsilon}{2} + r_{\shortparallel} \omega \right)^{2}
\right ], 
\end{multline}
where $\upsilon  = \omega - \omega'$ is the change in angular velocity,
leads to an exact expression for the jump moments.
In terms of parabolic cylinder functions, 
$\operatorname{D}_{n}(z) = \left( \exp[-z^{2}/4] / \Gamma[-n] \right) \int_{0}^{\infty} \exp[-zx-x^{2}/2] x^{-n-1} dx$ (for $n < 0$) the results are 
\begin{multline}
a_{n}(\omega) =
\frac{2^{n}}{\sqrt{2 \pi}}
\Gamma[n+2]
\sum_{i} \rho_{i}
\left(\frac{m}{k_{B} T_{i}}\right)^{-\frac{n+1}{2}}
\\
\times \oint dl_{i} \rho_{i}
\left( r_{\shortparallel} + \frac{I}{m r_{\shortparallel}}\right)^{-n}
\\
\times \exp \left[- \frac{m}{4 k_{B} T_{i}} r_{\shortparallel}^{2} \omega^{2} \right ]
\operatorname{D}_{-n-2} \left[\sqrt{\frac{m}{k_{B} T_{i}}} r_{\shortparallel} \omega \right]. 
\end{multline} 
Rescaling the jump moments using dimensionless variables $\xi  = \omega \sqrt{I /
k_{B} T_{\text{eff}}}$ and $\varepsilon = r_{0} \sqrt{m/I}$, where $r_{0}^{2} = I /M$, leads to
%\begin{widetext}
\begin{multline}
A_{n}(\xi ) = (\sqrt{I / k_{B} T_{\text{eff}}})^{n} a_{n}(\xi)
\\
 = \frac{2^{\frac{3n - 1}{2}}}{\sqrt{\pi}}
\sum_{i} \rho_{i}
\sqrt{\frac{k_{B} T_{i}}{m}}
\left(\frac{T_{i}}{T_{\text{eff}}} \right )^{n/ 2}
\\
\times \oint dl_{i}
\left(\frac{\varepsilon r_{\shortparallel} / r_{0}}{1 + \varepsilon^{2} 
(r_{\shortparallel} / r_{0})^{2}}\right )^{n}
\\
\times \exp \left[- \frac{\varepsilon^{2}}{2} \frac{T_{\text{eff}}}{T_{i}}
 (r_{\shortparallel} / r_{0})^{2}
 \xi ^{2}
\right ]\\
\times \Biggl(
\Gamma \left[\frac{n + 2}{2} \right ]
\Phi \left[\frac{n + 2}{2}; \frac{1}{2}; 
\frac{\varepsilon^{2}}{2} \frac{T_{\text{eff}}}{T_{i}}
 (r_{\shortparallel} / r_{0})^{2}
\xi ^{2} \right ]
\\
- \sqrt{2} \varepsilon  \sqrt{\frac{T_{\text{eff}}}{T_{i}}}
\frac{r_{\shortparallel}}{ r_{0}}  \xi \;
\Gamma \left[\frac{n + 3}{2} \right ]
\\
\times \Phi \left[\frac{n + 3}{2};
\frac{3}{2};
\frac{\varepsilon^{2}}{2} \frac{T_{\text{eff}}}{T_{i}}
(r_{\shortparallel} / r_{0})^{2}
 \xi ^{2} \right ]
\Biggr).
\label{eq:moments}
\end{multline}
%\end{widetext}
Here $\Phi$ represents Kummer's function \cite{gradshteyn}.

We can express both the exponential function and Kummer's function in a power series,
\begin{equation}
\exp[z] = 1 + \frac{z}{1!}
+ \frac{z^{2}}{2!} 
+\frac{z^{3}}{3!}
+ \cdots,
\end{equation}
\begin{multline}
\Phi[\alpha; \gamma; z] = \;_{1}F_{1} [\alpha; \gamma; z] = 1 + \frac{\alpha}{\gamma} \frac{z}{1!}
+ \frac{\alpha (\alpha + 1)}{\gamma (\gamma + 1)} \frac{z^{2}}{2!} 
\\
+ \frac{\alpha (\alpha + 1)(\alpha + 2)}{\gamma (\gamma + 1)(\gamma + 2)} \frac{z^{3}}{3!}
+ \cdots,
\end{multline}
Considering that the parameter $\varepsilon = \sqrt{m/M}$ is small for gas particles much lighter than the motor, we arrive at a series expansion for the jump moments in $\varepsilon$.
We substitute this expansion in the set of equations (Eq.\:\ref{eq:moments}) coupling the jump moments $A_{n}(\xi)$ with the moments of the angular velocity $\langle \xi^{n}  \rangle$.
For $n = 1$, with $\tau =
\varepsilon^{2} t$, this results in
\begin{align} 
\partial_{\tau} \langle \xi  \rangle &= \varepsilon^{-2} \langle A_{1}(\xi ) \rangle =
\sum_{i} \rho_{i} \sqrt{\frac{k_{B} T_{i}}{m}}
\nonumber\\
\times \biggl[
&\varepsilon^{-1} \sqrt{\frac{T_{i}}{T_{\text{eff}}}}  \oint dl_{i} \left(\frac{r_{\shortparallel}}{r_0}\right)
- 2 \sqrt{\frac{2}{\pi}} \langle \xi  \rangle  \oint dl_{i} \left(\frac{r_{\shortparallel}}{r_0}\right)^{2}
\nonumber\\
+ &\varepsilon \left(
\sqrt{\frac{T_{\text{eff}}}{T_{i}}} \langle \xi ^{2} \rangle
-\sqrt{\frac{T_{i}}{T_{\text{eff}}}}\right) \oint dl_{i} \left(\frac{r_{\shortparallel}}{r_0}\right)^{3}
\nonumber\\
+ &\frac{ \varepsilon^{2}}{3} \sqrt{\frac{2}{\pi}} \left(6 \langle \xi  \rangle -
\frac{T_{\text{eff}}}{T_{i}}\langle \xi ^{3}
\rangle \right)  \oint dl_{i} \left(\frac{r_{\shortparallel}}{r_0}\right)^{4}
\nonumber\\
+ &\varepsilon^{3} \left(\sqrt{\frac{T_{i}}{T_{\text{eff}}}} - 
\sqrt{\frac{T_{\text{eff}}}{T_{i}}}
\langle \xi ^{2}
\rangle \right)   \oint dl_{i} \left(\frac{r_{\shortparallel}}{r_0}\right)^{5} \biggr]
+ O(\varepsilon^{4}). \label{eq:epsilonexpansion}
\end{align}
The term in $\varepsilon^{-1}$ disappears because
\begin{equation}
\oint dl_{i} r_{\shortparallel} = \oint \vec{dl_{i}} \cdot \vec{r} = \int_{A_{i}} (\nabla \times \vec{r}) \cdot \vec{e_{z}} dA_{i} =0. \label{eq:closed}
\end{equation}
Similarly for $n=2$,
\begin{align} 
\partial_{\tau} \langle \xi^2\rangle &=\sum_i  \rho_i\sqrt{\frac{k_B T_i}{m}}
\nonumber\\
\times \biggl[-&4\sqrt{\frac{2}{\pi}}
\left(-\frac{T_i}{T_{\text{eff}}}+\langle \xi^2\rangle\right)\oint dl_{i} \left(\frac{r_{\shortparallel}}{r_0}\right)^2
\nonumber\\
-2&\varepsilon\left(4\sqrt{\frac{T_i}{T_{\text{eff}}}}\langle \xi\rangle-\sqrt{\frac{T_{\text{eff}}}{T_i}}\langle \xi^3\rangle\right)
\oint dl_{i} \left(\frac{r_{\shortparallel}}{r_0}\right)^3 
\nonumber\\
 +2&\varepsilon^2\sqrt{\frac{2}{\pi}}\left(-4\frac{T_i}{T_{\text{eff}}}+5\langle \xi^2\rangle-\frac{1}{3}\frac{T_{\text{eff}}}{T_i}\langle \xi^4\rangle
\right)\oint dl_{i} \left(\frac{r_{\shortparallel}}{r_0}\right)^4 
\nonumber\\
+2&\varepsilon^3\left(7\sqrt{\frac{T_i}{T_{\text{eff}}}}\langle \xi\rangle-2\sqrt{\frac{T_{\text{eff}}}{T_i}}\langle \xi^3\rangle\right)\oint dl_{i} \left(\frac{r_{\shortparallel}}{r_0}\right)^5 \biggr]
\nonumber\\
+O(&\varepsilon^4).
\end{align}

To lowest order in $\varepsilon$ we can extract from 
Eq.~(\ref{eq:epsilonexpansion}) a linear relaxation law for rotational movement, $I 
\partial_{t} \langle \omega
\rangle = \tau_{f}$, describing a net frictional torque $\tau_{f}$ exerted on the motor as a result of all collisions .With  $\tau_{f} = - \gamma \langle \omega \rangle$ and $\gamma = \sum_{i} \gamma_{i}$ we derive a microscopic expression for the friction coefficient $\gamma_{i}$ of each part of the object:
\begin{equation}
\gamma_{i} 
= 4 \rho_{i} \sqrt{\frac{k_{B} T_{i} m}{2 \pi}} \oint dl_{i} r_{\shortparallel}^{2}.
\label{eq:frictioncoefficient}
\end{equation}

To order $\varepsilon^{2}$ the average angular velocity in a stationary state  appears from Eq.~(\ref{eq:epsilonexpansion}) as
\begin{align}
\langle \omega \rangle
&= \frac{\sqrt{2 \pi k_{B} m}}{4I}
\frac{\sum_{i} \rho_{i} (T_{\text{eff}} - T_{i}) \oint dl_{i} r_{\shortparallel}^{3}}
{\sum_{i}\rho_{i} \sqrt{T_{i}}
\oint dl_{i} r_{\shortparallel}^{2}}
\nonumber\\
&= \frac{k_{B} m}{\gamma I} \sum_{i} \rho_{i} (T_{\text{eff}} - T_{i}) \oint dl_{i} r_{\shortparallel}^{3},
\label{eq:omega}
\end{align}
with second moment,
\begin{equation}
\langle \omega^{2} \rangle
= \frac{k_{B} T_{\text{eff}}}{I},
\end{equation}
while the effective temperature is found according to its definition,
\begin{equation}
T_{\text{eff}} = (\sum_i{\gamma_{i} T_i}) / (\sum_i{\gamma_{i}}).
\label{Teff}
\end{equation}
Using higher order terms in the expansions for $\langle \xi^{n}  \rangle$ results in correction terms to the expressions for $\langle \omega \rangle$ and $\langle \omega^{2} \rangle$. The second terms are in both a factor $m/M$ smaller than the first terms.

%%%%%%%%%%%%%%%%%%%%%%%%%%%%%%%%%%%%%%%%%%%%%%
\section{Three-dimensional model of the motor}
%%%%%%%%%%%%%%%%%%%%%%%%%%%%%%%%%%%%%%%%%%%%%%
\label{section:3dmotor}
The results of a fully three-dimensional analysis are very similar to those derived in the previous two-dimensional case. We will clarify the key differences here.
\begin{figure}
    \begin{center}
        \includegraphics[width=0.8\columnwidth]{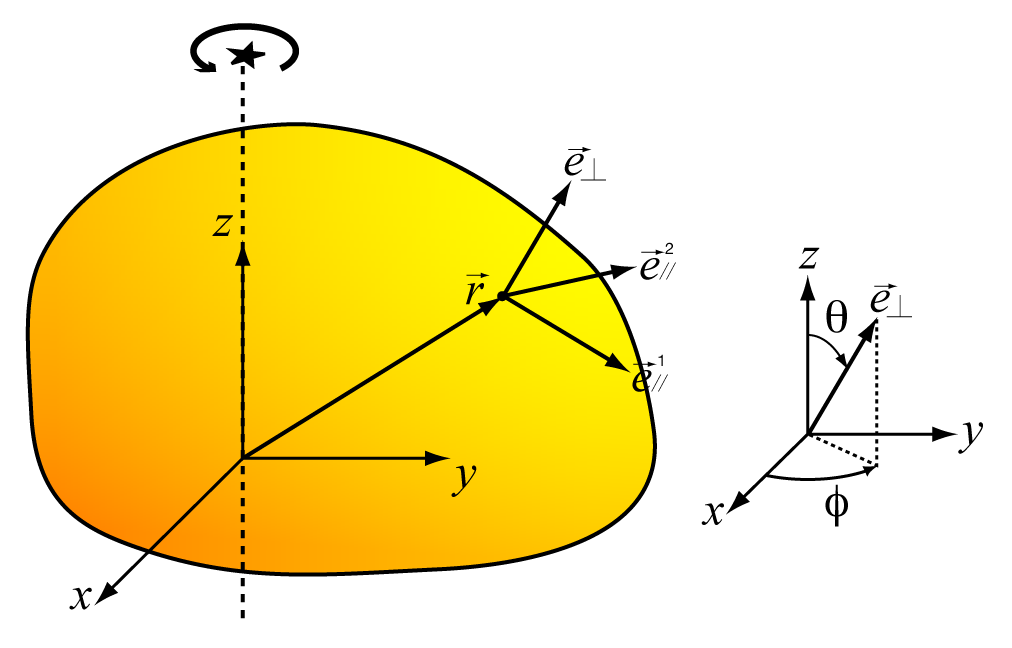}
    \end{center}
    \caption{In each reservoir, the motor part can be described by the boundary $\vec{r}(x,y)$, where the rotation axis is at the origin of the reference frame. It is convenient to also know explicitly the orientation of the boundary at any point. This is given by the polar angle $\varphi$ of the normal outward vector $\vec{e}_{\perp}$ on the surface at this point. The unit vector $\vec{e}_{\shortparallel}$ is tangential to the boundary.
    \label{fig:coordinates3d}}
\end{figure}

The motor parts are now determined by their surface $S_{i}$ ($i$ is the reservoir in which the part resides), to be described by a vector $\vec{r}(x,y,z)$ from the origin. Again we choose the rotation axis to be along the the $z$-axis.
The orientation of the surface at a certain location $(x,y,z)$ is uniquely determined by the normal
outward unit vector, $\vec{e}_{\perp} = (\sin \theta \cos
\varphi, \sin \theta \sin \varphi, \cos
\theta)$, described by two angles $\theta$ and $\varphi$, polar and azimuthal angles in a spherical coordinate system. 

The assumption that there is only a central force during an interaction with a gas particle, entails that there is conservation of momentum of a gas particle along any tangential direction. So for the speed of the gas particle before ( $\vec{v'} = (v'_{x}, v'_{y},
v'_{z})$) and after ($\vec{v} = (v_{x}, v_{y}, v_{z})$)  the collision, we can write this condition formally as
\begin{equation}
\vec{v'} \cdot \vec{e}_{k,\shortparallel}  = \vec{v} \cdot
\vec{e}_{k,\shortparallel}, \quad k = 1,2,
\end{equation}
where $\vec{e}_{1,\shortparallel}$ and $\vec{e}_{2,\shortparallel}$ are two distinct unit vectors perpendicular to $\vec{e}_{\perp}$. It is convenient to use
$\vec{e}_{1,\shortparallel} = (-\sin \varphi, \cos \varphi, 0)$ and
$\vec{e}_{2,\shortparallel} = (\cos \theta \cos \varphi, \cos
\theta \sin \varphi, -\sin \theta)$.
Together with conservation of total energy and angular momentum in the $z$-direction (the expressions are the same as in the two-dimensional analysis) we find a relation for the change of angular velocity induced by one collision:
 \begin{equation}
\omega = \omega' + 2 \frac{(\vec{V'} - \vec{v'}) \cdot \vec{e}_{\perp}}{r_{\shortparallel} + \frac{I}{m r_{\shortparallel}}}.
\end{equation}
$(\vec{V'} - \vec{v'}) \cdot \vec{e}_{\perp}$ is the component of the velocity difference between motor and gas particle perpendicular to the surface at the place of impact.
$r_{\shortparallel}$ is now defined as
 \begin{equation}
r_{\shortparallel} = - x \sin \theta \sin \varphi + y \sin \theta \cos \varphi
= \sin \theta \:\vec{r} \cdot \vec{e}_{1,\shortparallel}.
\end{equation}
$r_{\shortparallel}$ is zero in locations where the surface is perpendicular to $\vec{r}$, these coincide with zero momentum transfer. Maximal momentum transfer and $r_{\shortparallel}$ occurs when the tangential plane to the surface at this location crosses the rotation axis.

The transition probability $dW(\omega | \omega')$ caused by all possible collisions with a surface element $dS_{i}$ of the motor, is then found by integrating over all velocities that obey the collision rules,
\begin{multline}
dW_{i}(\omega | \omega') = dS_{i}
\int_{-\infty}^{+\infty} \negthickspace \negthickspace dv'_{x}
\int_{-\infty}^{+\infty} \negthickspace \negthickspace dv'_{y}
\int_{-\infty}^{+\infty} \negthickspace \negthickspace dv'_{z}
\\
\times H[(\vec{V'} - \vec{v'}) \cdot \vec{e}_{\perp}]
 |(\vec{V'} - \vec{v'}) \cdot \vec{e}_{\perp}|
 \\
\times \rho_{i} \phi_{i}(v'_{x}, v'_{y}, v'_{z}) \delta \left[\omega - \omega' -  2 \frac{(\vec{V'} - \vec{v'}) \cdot \vec{e}_{\perp}}{r_{\shortparallel} + \frac{I}{m r_{\shortparallel}}}
\right].
\end{multline}
Adding the contributions of all surface elements $dS_{i}$ in all reservoirs $i$, gives us the total transition probability,
\begin{equation}
 W(\omega | \omega') = \sum_{i} \int_{\text{surface}}\negthickspace \negthickspace \negthickspace \negthickspace \negthickspace  \negthickspace   \negthickspace \negthickspace dW_{i}(\omega | \omega').
\end{equation}

Again, for a Maxwellian velocity distribution,
\begin{equation}
\phi_{i}(v_{x}, v_{y}, v_{z}) = \left(\frac{m}{2 \pi k_{B} T_{i}}\right
)^{3/2}\exp \left(\frac{- m (v_{x}^2 + v_{y}^2 + v_{z}^2)}{2 k_{B} T_{i}}\right
),
\end{equation}
we can do the integration over $v_{x}, v_{y}, v_{z}$ analytically and find
\begin{multline} W(\omega | \omega') = \frac{1}{4} \sum_{i}
\int dS_{i} \rho_{i}
\sqrt{\frac{m}{2 \pi k_{B} T_{i}}}
\\
\times ( r_{\shortparallel} + \frac{I}{m r_{\shortparallel}})^{2}
H[(\omega - \omega') r_{\shortparallel}] |\omega -
\omega'|
\\
\times \exp \left[- \frac{m}{2 k_{B} T_{i}}
 \left((r_{\shortparallel} + \frac{I}{m r_{\shortparallel}}) \frac{\omega' - \omega}{2} - r_{\shortparallel} \omega' \right)^{2}
\right].
\end{multline}
This expression is identical to its two-dimensional equivalent (Eq.\:\ref{transitionprobability}), apart from the different definition of $r_{\shortparallel}$, and obviously an integration over the surface instead of the boundary.
The previous algebraic technique can then be applied to derive results for a general shape of the motor, such as for the average angular velocity in a steady state,
\begin{equation}
\langle \omega \rangle
= \frac{\sqrt{2 \pi k_{B} m}}{4I}
\frac{\sum_{i} \rho_{i} (T_{\text{eff}} - T_{i}) \int dS_{i} r_{\shortparallel}^{3}}
{\sum_{i}\rho_{i} \sqrt{T_{i}}
\int dS_{i} r_{\shortparallel}^{2}},
\end{equation}
and the friction coefficient,
\begin{equation}
\gamma 
=  \sum_{i} \gamma_{i}
= \sum_{i} 4 \rho_{i} \sqrt{\frac{k_{B} T_{i} m}{2 \pi}} \int dS_{i} r_{\shortparallel}^{2}.
\end{equation}
where $T_{\text{eff}}$ is still defined as
\begin{equation}
T_{\text{eff}} = (\sum_i{\gamma_{R, i} T_i}) / (\sum_i{\gamma_{R, i}}).
\end{equation}

%%%%%%%%%%%%%%%%%%%%%%%%%%%%%%%%%%%%%%%%%%%%%%
\section{Analysis and discussion}
%%%%%%%%%%%%%%%%%%%%%%%%%%%%%%%%%%%%%%%%%%%%%%
Now that we derived analytical results for any shape and any number of reservoirs, we are ready to analyze concrete systems. We are interested in the role of external parameters, such as the temperature and the density of the gas, and in the construction and shape of the motor itself. Much of the analysis can be applied to the simpler two-dimensional case, but references to the three-dimensional case are made where they are appropriate. 

%%%%%%%%%%%%%%%%%%%%%%%%%%%%%%%%%%%%%%%%%%%%%%
\subsection{Temperature gradient}
%%%%%%%%%%%%%%%%%%%%%%%%%%%%%%%%%%%%%%%%%%%%%%
When the thermal reservoirs are at equilibrium with each other, we immediately see from Eq.\:[\ref{Teff}] that $T_{1} = T_{2} = \cdots = T_{i} = T_{\text{eff}}$, independent of the construction we propose. The average angular velocity
\begin{equation}
\langle \omega \rangle
= \frac{k_{B} m}{\gamma I} \sum_{i} \rho_{i} (T_{\text{eff}} - T_{i}) \oint dl_{i} r_{\shortparallel}^{3},
\label{omegarevisited}
\end{equation}
becomes zero. It is impossible to extract net motion from a system in equilibrium.

%%%%%%%%%%%%%%%%%%%%%%%%%%%%%%%%%%%%%%%%%%%%%%
\subsection{Chirality}
%%%%%%%%%%%%%%%%%%%%%%%%%%%%%%%%%%%%%%%%%%%%%%
The next element we want to discuss is the factor $\oint dl_{i} r_{\shortparallel}^{3}$ in Eq.\:[\ref{omegarevisited}].  Consider a motor shape in one reservoir $i$ that is symmetrical with respect to a plane through the rotation axis. A simple argument reveals that $\oint dl_{i} r_{\shortparallel}^{3} = 0$: for every point $(x,y)$ on the boundary of the shape with value $r_{\shortparallel}$ there can be found a point $(x',y')$ for which $r'_{\shortparallel} = - r_{\shortparallel}$. The contour integral of $r_{\shortparallel}$ is therefore zero, considering that the line element $dl_{i}$ is positive.

A construction that consists entirely of symmetric shapes will yield zero average rotation. Such a construction in its most simple form could consist of flat blades through the rotation axis in every reservoir. To find a net angular velocity, the motor must have at least one chiral part. The factor $\oint dl_{i} r_{\shortparallel}^{3}$ will be analyzed in more detail in a later section, and we will show that under certain conditions it can also become zero even for a chiral configuration.

%%%%%%%%%%%%%%%%%%%%%%%%%%%%%%%%%%%%%%%%%%%%%%
\subsection{Friction and propulsion}
%%%%%%%%%%%%%%%%%%%%%%%%%%%%%%%%%%%%%%%%%%%%%%
In the full expression
\begin{equation}
\langle \omega \rangle
= \frac{\sqrt{2 \pi k_{B} m}}{4I}
\frac{\sum_{i} \rho_{i} (T_{\text{eff}} - T_{i}) \oint dl_{i} r_{\shortparallel}^{3}}
{\sum_{i}\rho_{i} \sqrt{T_{i}}
\oint dl_{i} r_{\shortparallel}^{2}},
\label{omegafullrevisited}
\end{equation}
the factor $\oint dl_{i} r_{\shortparallel}^{2}$ in the denominator stems from the friction each motor part encounters while rotating in the gas. If we look at optimizing the motor, the first idea would be to minimize this factor. A surface where $r_{\shortparallel}$ is zero at every point corresponds to a sphere, but 
$\oint dl_{i} r_{\shortparallel}^{3}$ will be zero as well, resulting in zero net motion. Large average angular velocities will be obtained then by a compromise between a small $\oint dl_{i} r_{\shortparallel}^{2}$, and a large $\oint dl_{i} r_{\shortparallel}^{3}$. The propulsion of the motor originates in the factor  $\oint dl_{i} r_{\shortparallel}^{3}$. The largest friction will be experienced by shapes where $r_{\shortparallel}$ is maximal. This corresponds to a (flat) surface, or blade, through the rotation axis.

%%%%%%%%%%%%%%%%%%%%%%%%%%%%%%%%%%%%%%%%%%%%%%
\subsection{Motor configurations}
%%%%%%%%%%%%%%%%%%%%%%%%%%%%%%%%%%%%%%%%%%%%%%
\label{section: configurations}
We turn to the question of how to configure the motor. Leaving the exact choice of the shape for later, we tackle the following the question: if we have a certain part of the motor in one reservoir, how will the placement of the part in the other reservoir effect the motion of the motor
We start by proposing three simple constructions (see Fig.\:\ref{fig:3configurations}):
\begin{enumerate}
\item The shapes are identical in both reservoirs (Fig.\:\ref{fig:3configurations}a). This includes the location of the rotation axis with respect to the shape. The exact shape can be determined afterwards.
With
\begin{alignat}{3}
 \oint dl_{1} r_{\shortparallel}^{2} &=   \oint dl_{2} r_{\shortparallel}^{2} &=   \oint dl r_{\shortparallel}^{2},\\
 \oint dl_{1} r_{\shortparallel}^{3} &=   \oint dl_{2} r_{\shortparallel}^{3} &=   \oint dl r_{\shortparallel}^{3},
\end{alignat}
Eq.\:[\ref{omegafullrevisited}] simplifies to
\begin{equation}
\langle \omega \rangle
= \frac{\sqrt{2 \pi k_{B} m}}{4I}
\frac{\rho_{1} \rho_{2} (T_{2}^{1/2} - T_{1}^{1/2})(T_{2} - T_{1})}
{(\rho_{1} T_{1}^{1/2} + \rho_{2} T_{2}^{1/2})^{2}}
\frac{\oint dl\,  r_{\shortparallel}^{3}}{\oint  dl\,  r_{\shortparallel}^{2}}.
\label{eq:omegaequalparts}
\end{equation}
\begin{figure}
    \begin{center}
        \includegraphics[width=\columnwidth]{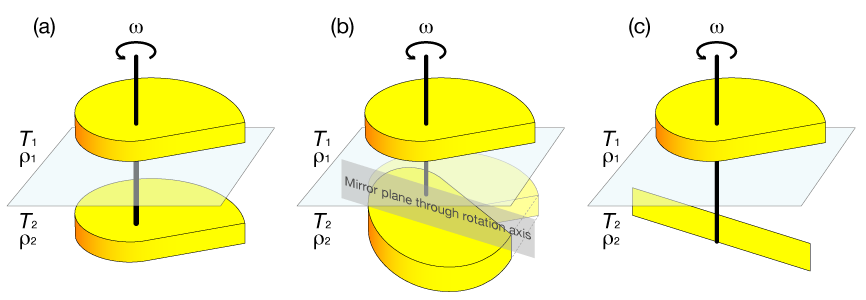}
    \end{center}
    \caption{Three possible configurations of the motor parts in a two-reservoir system. In (a) the part in the first reservoir is copied exactly to the other reservoir. Not only the shapes are identical, also their position relative to the rotation axis. In (b) the motor part in one reservoir is reflected in the other reservoir. The reflecting plane passes through the rotation axis. In (c) a general shape is combined with a blade: a plane of length $L$ (and height $H$ in three dimensions). Note that the system is rotationally invariant in each reservoir separately.
    \label{fig:3configurations}}
\end{figure}

\item The shapes are still general but they are exact mirror images of each other in both reservoirs (Fig.\:\ref{fig:3configurations}b). Also the location of the rotation axis with respect to the shape is mirrored. The mirror axis (plane) is through the rotation axis, but its orientation is of no importance, as our system is rotationally invariant.
Writing
\begin{alignat}{3}
 \oint dl_{1} r_{\shortparallel}^{2} &=   \oint dl_{2} r_{\shortparallel}^{2} &=   \oint dl r_{\shortparallel}^{2},\\
 \oint dl_{1} r_{\shortparallel}^{3} &=   -\oint dl_{2} r_{\shortparallel}^{3} &=   \oint dl r_{\shortparallel}^{3},
\end{alignat}
Eq.\:[\ref{omegafullrevisited}] now becomes
\begin{equation}
\langle \omega \rangle
= \frac{\sqrt{2 \pi k_{B} m}}{4I}
\frac{\rho_{1} \rho_{2} (T_{2}^{1/2} + T_{1}^{1/2})(T_{2} - T_{1})}
{(\rho_{1} T_{1}^{1/2} + \rho_{2} T_{2}^{1/2})^{2}}
\frac{\oint dl\,  r_{\shortparallel}^{3}}{\oint  dl\,  r_{\shortparallel}^{2}}.
\label{eq:omegamirrorparts}
\end{equation}

\item We use a general (yet unknown) shape in the first reservoir, while in the second reservoir we put a blade of length $L$, rotating about one end (Fig.\:\ref{fig:3configurations}c). Omitting the index $i = 1$, and identifying
\begin{align}
\oint dl_{2} r_{\shortparallel}^{2} &=   2 L^{3}/3,\\
 \oint dl_{2} r_{\shortparallel}^{3} &=  0,
\end{align}
we obtain
\begin{equation}
\langle \omega \rangle
= \frac{\sqrt{2 \pi k_{B} m}}{4I}
\frac{\rho_{1} \rho_{2} T_{2}^{1/2}(T_{2} - T_{1}) (2 L^{3}/3) \oint dl\,  r_{\shortparallel}^{3}}
{(\rho_{1} T_{1}^{1/2} \oint  dl\,  r_{\shortparallel}^{2}
 + \rho_{2} T_{2}^{1/2} (2 L^{3}/3))^{2}}.
\label{eq:shapeandblade}
\end{equation}
\end{enumerate}
Comparing the three suggested configurations, we see configuration (1) is even when the temperature difference $\Delta T = T_{1} - T_{2}$ is inverted, while (2) and (3) are odd. For small temperature differences, $\langle \omega \rangle$ is approximately parabolic in $\Delta T$, while (2) and (3) are linear. For small $\Delta T$ therefore (1) yields much lower angular speeds than (2) and (3).

For a small temperature difference a rather technical calculation shows that the average angular velocity for configuration (2) is at least twice that of construction (3) for the same general shape with similar linear dimensions as the blade. 

In a numerical procedure described later, we discovered that the configuration of two mirror shapes in the two reservoirs produces the maximal average angular velocity. In other words, given a certain part in one reservoir, the highest average angular velocity is obtained by using the reflected shape in the other reservoir.

\begin{figure}
    \begin{center}
        \includegraphics[width=0.9\columnwidth]{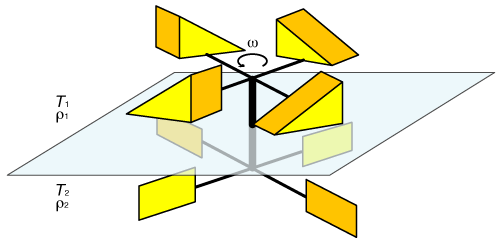}
    \end{center}
    \caption{A motor with multiple identical shapes in each reservoir. For the depicted configuration with equal multiples, $n = m =4$, we show that the resulting average angular velocity is the same as that for only one shape in each reservoir.
    \label{fig:multipleblades}}
\end{figure}
One could think of multiple but identical structures (blades) in each reservoir, as illustrated in Fig.\:\ref{fig:multipleblades}. Ignoring the increased probability of multiple collisions of gas particles with the motor, our theory leads to an average angular velocity
\begin{widetext}
\begin{equation}
\langle \omega \rangle
= \frac{\sqrt{2 \pi k_{B} m}}{4I}
\frac{\rho_{1} \rho_{2} (T_{2} - T_{1})
\left(T_{2}^{1/2} \oint dl_{1}\,  r_{\shortparallel}^{3} \oint dl_{2}\,  r_{\shortparallel}^{2}
- T_{1}^{1/2} \oint dl_{2}\,  r_{\shortparallel}^{3} \oint dl_{1}\,  r_{\shortparallel}^{2}\right)}
{\left(\sqrt{\frac{n}{m}}\rho_{1} T_{1}^{1/2} \oint dl_{1}\,  r_{\shortparallel}^{2}
+ \sqrt{\frac{m}{n}} \rho_{2} T_{2}^{1/2}\oint dl_{2}\,  r_{\shortparallel}^{2}\right)^{2}},
\end{equation}
\end{widetext}
for a system with $n$ identical blades in reservoir 1 and $m$ identical blades in reservoir 2.
The appearing contour integrals are over \emph{one} shape of the set of identical shapes. For the same number of shapes in both reservoirs, $n = m$, the average angular velocity is the same as with only one blade in each reservoir. The result also shows that it is beneficial to have the highest number of blades in the reservoir with the highest $\rho T^{1/2} \oint dl\,  r_{\shortparallel}^{2}$ factor, or simply the highest $\rho T^{1/2}$ factor if the blades have the same shape in both reservoirs.

%%%%%%%%%%%%%%%%%%%%%%%%%%%%%%%%%%%%%%%%%%%%%%
\subsection{Globular proteins}
%%%%%%%%%%%%%%%%%%%%%%%%%%%%%%%%%%%%%%%%%%%%%%
\label{globularproteins}
Looking for real-world candidates to fill the role of our Brownian motor, we turn our attention to biological systems. In the further analysis we want to use physical values for the dimensions, masses and so on. A possibility is to apply our model to globular proteins, which could give the shape of the motor parts. The two parts would reside in a water environment, separated by a lipid membrane. 

To obtain orders of magnitude for our results we will refer often to the values in Table\:\ref{table:globularprotein}.

\begingroup
\squeezetable
\begin{table}[h]
\begin{center}
\begin{ruledtabular}
\begin{tabular}{l l l }
Mass of one part & $M/2$ & $1.66\times10^{-22}  \text{kg}$\\
Density of the motor & $\rho_{m}$ &  $1380 \:\text{kg m}^{-3}$\\
Volume of one part & $V$ &  120 $\text{nm}^{3}$\\
Radius of one part (if assumed spherical) & $R$ &  3 nm\\
Particle mass ($\text{H}_{2}\text{O}$) & $m$ &  $2.992\times10^{-26}  \text{kg}$\\
Reservoir temperature & $T_{1}, T_{2}$ & $\pm 300$ K \\
Reservoir particle density & $\rho_{1}, \rho_{2}$ & $\pm 3.3\times 10^{28} \text{m}^{-3}$
\end{tabular}
\end{ruledtabular}
\caption{\label{table:globularprotein}
Typical parameters used for the motor and environment. The values of the individual parts correspond to those of globular proteins.}
\end{center}
\end{table}
\endgroup

%%%%%%%%%%%%%%%%%%%%%%%%%%%%%%%%%%%%%%%%%%%%%%
\subsection{External parameters}
%%%%%%%%%%%%%%%%%%%%%%%%%%%%%%%%%%%%%%%%%%%%%%
\begin{figure}
    \begin{center}
        \includegraphics[width=\columnwidth]{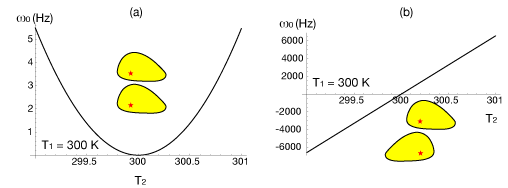}
    \end{center}
    \caption{The temperature dependency of $\omega_{0}$ for a configuration of (a) identical shapes and of (b) mirror shapes in the two reservoirs. One reservoir is kept on a fixed temperature $T_{1} = 300 \text{K}$, while the temperature $T_{2}$ in the other reservoir changes in a range of $300 \pm 1 \text{K}$. 
For (a) we see a nearly parabolic dependency on $T_{2} - T_{1}$. The sense of rotation remains the same if we switch part 1 and part 2 of the motor, and the angular velocity is rather small.
For (b) we see a much larger effect, a nearly linear dependency on $T_{2} - T_{1}$, and the sense of rotation is inverted by switching part 1 and part 2 of the motor. The densities are taken the same in both reservoirs.
   \label{fig:omega0}}
\end{figure}
In the cases of identical shapes and mirror shapes (Eqs.\:[\ref{eq:omegaequalparts}, \ref{eq:omegamirrorparts}]) we can separate from the expressions for $\langle \omega \rangle$ a shape-dependent factor,
\begin{equation}
\mathcal{S} =
\sqrt{A}
\frac{M}{I}
\frac{\oint dl\,  r_{\shortparallel}^{3}}{\oint  dl\,  r_{\shortparallel}^{2}}.
\end{equation}
A factor $\sqrt{A}$ (or $V^{1/3}$ if we prefer to work in three dimensions), which stands for the typical dimensions of a motor part, is multiplied to make $\mathcal{S}$ scale-invariant.
$\mathcal{S}$ will be discussed in detail in the next section.
Eqs.\:[\ref{eq:omegaequalparts}, \ref{eq:omegamirrorparts}] can then be written as
\begin{equation}
\langle \omega \rangle
= 
\omega_{0} \mathcal{S}.
\end{equation}
What remains is a factor $\omega_{0}$ that depends on the specific configuration, the reservoirs temperatures and densities, and the masses of the motor and particles:
\begin{equation}
\omega_{0} = \omega_{0}^{\pm}  = \frac{\sqrt{2 \pi k_{B} m}}{4 M \sqrt{A}}
\frac{\rho_{1} \rho_{2} (T_{2}^{1/2} \mp T_{1}^{1/2})(T_{2} - T_{1})}
{(\rho_{1} T_{1}^{1/2} + \rho_{2} T_{2}^{1/2})^{2}}.
\end{equation}
We have used the notation $\omega_{0}^{+}$ for the configuration with two identical shapes in the two reservoirs and $\omega_{0}^{-}$ for the configuration where the shapes are mirror images.

$\omega_{0}$ is also dependent of the size of the motor. Because the mass $M$ of the motor is also size-dependent, the full dependency could be written as $M \sqrt{A} = 2 \rho_{m} A^{3/2}$ in two dimensions, or $M V^{1/3} = 2 \rho_{m} V^{4/3}$ in three dimensions, if the density $\rho_{m}$ of the motor interior is considered constant.
Therefore $\omega_{0} \propto M^{-3/2}$ in two dimensions and  $\omega_{0} \propto M^{-4/3}$ in three dimensions.

Fig.\:\ref{fig:omega0} shows the temperature dependency of $\omega_{0}$, calculated with the physical values of Table\:\ref{table:globularprotein} and equal reservoir densities. As mentioned before $\omega_{0}^{-}$ is linear in $T_{2}-T_{1}$, while $\omega_{0}^{+}$ is quadratic. 

%%%%%%%%%%%%%%%%%%%%%%%%%%%%%%%%%%%%%%%%%%%%%%
\subsection{Shape factor}
%%%%%%%%%%%%%%%%%%%%%%%%%%%%%%%%%%%%%%%%%%%%%%
\label{section:shape}
Next, we consider the size-independent geometrical factor,
\begin{equation}
\mathcal{S} =
\sqrt{A}
\frac{M}{I}
\frac{\oint dl\,  r_{\shortparallel}^{3}}{\oint  dl\,  r_{\shortparallel}^{2}},
\end{equation}
which is comprised of an interior factor $M/I$, and an exterior (boundary) factor $\oint dl\,  r_{\shortparallel}^{3} / \oint  dl\,  r_{\shortparallel}^{2}$

The factor $M/I$ is actually independent of the mass of the motor because the inertial moment $I$ is proportional to the mass $M$. It only describes the spacial distribution of mass. For a homogeneous motor interior it is given by
\begin{equation}
\frac{M}{I} = \frac{\sum_{i} A_{i}}{\sum_{i} \int  r^{2} dA_{i}}.
\end{equation}
The integral is over the entire interior of the motor, and $r$ is the distance of an interior point to the rotation axis.

Finally the factor $\oint dl\,  r_{\shortparallel}^{3} / \oint  dl\,  r_{\shortparallel}^{2}$
depends on the exact form of the boundary of the motor parts, where $r_{\shortparallel}$ is to be measured from the location of the rotation axis. The integrals are over the entire boundary.

\begin{figure}
    \begin{center}
        \includegraphics[width=0.9\columnwidth]{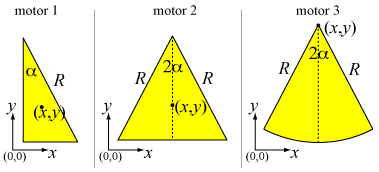}
    \end{center}
    \caption{Three simple shapes for the motor parts. All all determined by a minimal number of parameters. The shape is determined by the angle $\alpha$, the size by one side $R$, and the position with respect to the rotation axis (at the origin) by one coordinate $(x, y)$. For the triangles (Motor 1: right triangle; Motor 2: isosceles triangle) $(x, y)$ is the location of the center of mass, while for the disk sector (Motor 3) $(x, y)$ corresponds to the center of the disk.
    \label{fig:2dmotors}}
\end{figure}

To enable us to get an understanding of the geometrical factor,
we introduce three simple realizations (Fig.\:\ref{fig:2dmotors}) (in two dimensions and with homogeneous mass distributions), of which the boundary can easily be described analytically. For these prototype shapes all factors can be expressed in closed form.
The shape of the motors parts are respectively a right triangle (Motor 1), an isosceles triangle (Motor 2) and a disk sector (Motor 3). 
Both the dimensions of the motor parts ($R$) and the shape ($\alpha$) are fixed with one parameter, making a comparative study easier.
We would also like to specify the location of the motor part relative to the rotation axis with one representative point with coordinate $(x, y)$.
For motors 1 and 2 we choose the center of mass and for motor 3 the center of the disk sector (see (Fig.\:\ref{fig:2dmotors})) as this representative point .
For these simple motors we can calculate analytical expressions for $\oint  dl\,  r_{\shortparallel}^{2}$ and  $\oint  dl\,  r_{\shortparallel}^{3}$. 
As these expressions are rather long, and the details are not of immediate relevance, we have put them in the appendix. As an example consider the Motor 2 case:
\begin{multline}
\oint  dl\,  r_{\shortparallel}^{2} = 
\frac{2 R}{9}  \biggl(R^2+3 R y \cos 3 \alpha +9 y^2 + 9 x^2 \sin \alpha  
\\
+\sin ^2 \alpha 
   \left(R^2 (3 \sin \alpha -2 \cos 2 \alpha )+ 9 x^2 -9 y^2\right) \biggr),
\end{multline}
and
\begin{multline}
\oint  dl\,  r_{\shortparallel}^{3} = 
\frac{R}{3}  x \sin 2 \alpha
\biggl(R (3 y+R \cos \alpha ) (1 - 2 \cos 2 \alpha )
\\
+3 \left(x^2-3 y^2\right) \cos \alpha \biggr).
\label{motor2s3}
\end{multline}
The intertial moments $I$ for the three motors are given in Table\:\ref{table:inertialmoments}.
\begin{table}
\begin{center}
\begin{ruledtabular}
\begin{tabular}{l l}
Motor 1 & $I / M = R^{2}/18 + x^{2} + y^{2}$ \\
Motor 2 & $I / M = R^2 (2 - \cos 2 \alpha) / 18 +x^2+y^2$ \\
Motor 3 & $I / M = \frac{R^2}{2}-\frac{4 R y \sin \alpha  }{3 \alpha }+x^2+y^2$
\end{tabular}
\end{ruledtabular}
\caption{The ratio of the inertial moment $I$ over the mass $M$ of the motors in Fig.\:\ref{fig:2dmotors} given as a function of the location $(x, y)$ (of the center of mass for Motors 1 and 2, and of the center of the disk for Motor 3) with respect to the rotation axis and the shape parameters, angle $\alpha$ and size $R$. The distribution of mass within the motors is assumed homogeneous.
\label{table:inertialmoments}}
\end{center}
\end{table}
Some physical properties are immediately apparent from these expressions. For example $\oint  dl\,  r_{\shortparallel}^{3}$, and hence the angular velocity $\langle \omega \rangle$, is zero when
\begin{itemize}
\item $x = 0$: this is when the rotation axis is on the symmetry axis of the motor; there is no preferred sense of rotation,
\item $\sin 2 \alpha = 0$ or $\alpha = 0$ or $\alpha = \pi/2$: the motor is bar shaped, and loses its asymmetry (or chirality).
\end{itemize}
Note that $\oint  dl\,  r_{\shortparallel}^{2}$, which also appears as a factor in the expression for the friction coefficient, is \emph{not} zero if the shape is bar shaped (or symmtrical in general).

In general the $\oint  dl\,  r_{\shortparallel}^{3}$ factor (Eq.\:\ref{motor2s3} for Motor 2) describes the asymmetry of the motor. It also determines the sense of rotation. For example for Motor 2, the rotation sense is inverted when the rotation axis is placed on opposite sides of the symmetry axis, $x = x_{0}$ and $x = -x_{0}$.

\begin{figure}
    \begin{center}
        \includegraphics[width=\columnwidth]{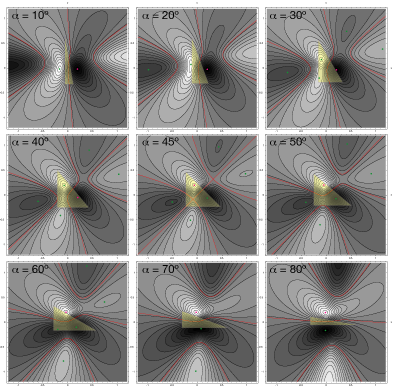}
    \end{center}
    \caption{Contour plots of the average angular velocity $\langle \omega \rangle(x,y)$ of Motor 1 as a function of the location of the rotation axis $(x, y)$, for several values of the angle $\alpha$. The center of mass of the motor is in the origin $(0,0)$.
    Regions in black (and white) correspond to locations for the rotation axis that yield the highest $\langle \omega \rangle$ (but in opposite sense). If the rotation axis is put on a red curve there is zero average rotation. Maxima in $\langle \omega \rangle(x,y)$ are marked by dots, the purple dot reveals the optimal place for the rotation axis (the one that gives the highest angular velocity). Note that one red curve for the $\alpha = 45^\circ$ realisation corresponds with a symmetry axis of the shape. The motor will not show directed motion if the axis is place there. Note also that mirror shapes  (such as $\alpha = 20^\circ$  and $\alpha = 70^\circ$) show opposite rotation sense, for all locations $(x, y)$.
   \label{fig:contoursmotor1}}
\end{figure}

\begin{figure}
    \begin{center}
        \includegraphics[width=\columnwidth]{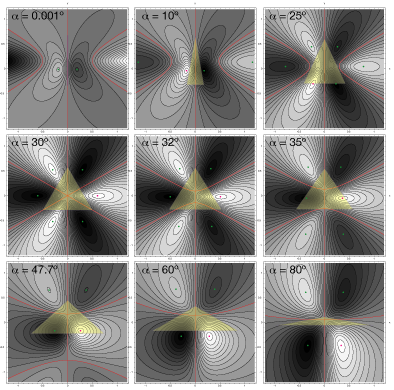}
    \end{center}
    \caption{Contour plots of the average angular velocity $\langle \omega \rangle(x,y)$ of Motor 2 as a function of the location of the rotation axis $(x, y)$, for several values of the angle $\alpha$. For a technical explanation see Fig.\:\ref{fig:contoursmotor1}. The symmetry of the shape is reflected in the $\langle \omega \rangle(x,y)$ plot, in particular in the locations for the rotation axis that correspond to zero average angular velocity (red curves): the $y$-axis for all the configurations and three symmetry axes for $\alpha = 30^\circ$. Note that the rotation sense is opposite for locations on opposite sides of a symmetry axis.
   \label{fig:contoursmotor2}}
\end{figure}

\begin{figure}
    \begin{center}
        \includegraphics[width=\columnwidth]{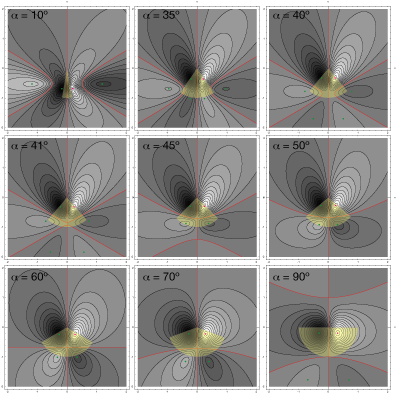}
    \end{center}
    \caption{Contour plots of the average angular velocity $\langle \omega \rangle(x,y)$ of Motor 3 as a function of the location of the rotation axis $(x, y)$, for several values of the angle $\alpha$. For a technical explanation see Fig.\:\ref{fig:contoursmotor1}.
   \label{fig:contoursmotor3}}
\end{figure}

More features can be seen from plots of the angular velocity of the motors  as a function of their shape and configuration, see Figs.\:\ref{fig:contoursmotor1} (Motor 1), \ref{fig:contoursmotor2} (Motor 2) and \ref{fig:contoursmotor3} (Motor 3). Note that a coordinate change was made, $x \rightarrow -x, y \rightarrow -y$. This means the coordinate $(x,y)$ in the plot corresponds to the location of the \emph{rotation axis} with respect to the representative point of the motor (center of mass or center of the disk), which is put in the origin of the plots. Figs.\:\ref{fig:contoursmotor1}, \ref{fig:contoursmotor2}, \ref{fig:contoursmotor3} of $\langle \omega \rangle(x,y)$ show lines of equal angular velocity (in black) and lines of zero average angular velocity (in red). Highest angular velocities are found in the black and white regions (but with opposite rotation sense). Local extrema of $\langle \omega \rangle(x,y)$ are represented by a green dot while a purple dot is the optimal location of the rotation axis. The shape of the motor is drawn in yellow.

We see that the red  curves that  signify zero average rotation can be straight lines when they correspond to a symmetry axis of the shape (the $y$-axis in Figs.\:\ref{fig:contoursmotor2}, \ref{fig:contoursmotor3} for all shapes $\alpha$, but also in Fig.\:\ref{fig:contoursmotor1} for $\alpha = 45^\circ$ for example), but in general they follow a curved path. The regions of opposite rotation sense, separated from each other by the red curves, form not so trivial patterns.

Also interesting to note is that the location of the rotation axis that yields the highest rotation speed is always in the vicinity of the rotating Brownian motor and often in its interior. Remember that the geometrical results are scale-invariant, and the relative locations of the maxima (and zero lines) are independent of the dimensions of the motor. 

\begin{figure}
    \begin{center}
        \includegraphics[width=0.7\columnwidth]{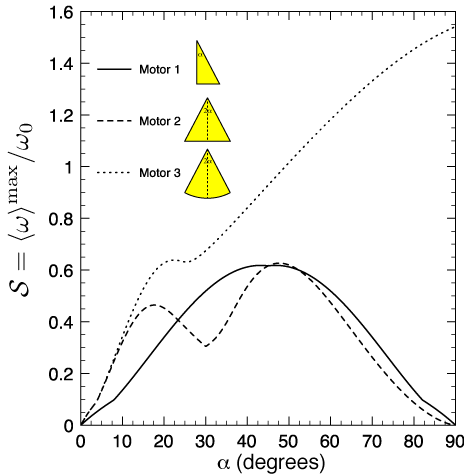}
    \end{center}
    \caption{For the simple motor types (Fig.\:\ref{fig:2dmotors}), we determined the position of the rotation axis that maximizes the average rotation speed for every shape, given by angle $\alpha$. The values of  $\alpha$ for which the shape factor $\mathcal{S}$ reaches a maximum are given in Table\:\ref{table:shapefactorsimplemotors}. It becomes clear that the shape is a key factor in the operation of the motor.
    \label{fig:omegaalpha}}
\end{figure}

\begin{figure}
    \begin{center}
        \includegraphics[width=\columnwidth]{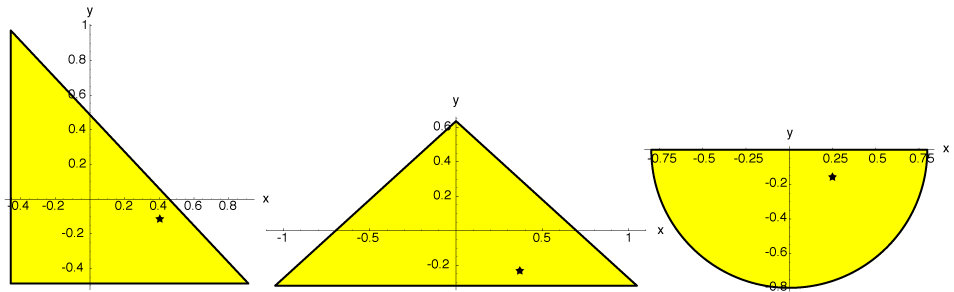}
    \end{center}
    \caption{The choices for the shape and location of the rotation axis for each of the three motor realizations that result in the highest average angular velocity of the motor. The location of the rotation axis is marked by a star.
    Although the initial constraints for Motor 1 and 2 are different (respectively the shape of a right and isosceles triangle), in their optimal configuration they are very similar. Motor 3 is optimal in the shape of a semi-disk, while we excluded the possibility of a concave shape.
    \label{fig:motorsshapemax}}
\end{figure}

For certain choices of the shape and especially of the location of the rotation axis the average rotation speed can become zero. It is therefore sensible to investigate which configurations yield the highest rotation speed. 
For the three simple motor realizations, we determine the location of the rotation axis that yields the highest shape factor $\mathcal{S}$ for every value of the shape parameter $\alpha$, as shown in Fig.\:\ref{fig:omegaalpha}.
The angle $\alpha$ that results in the highest $\mathcal{S}$ is listed in Table\:\ref{table:shapefactorsimplemotors} for each of the motors. The corresponding shapes are depicted in Fig.\:\ref{fig:motorsshapemax}.
\begin{table}
\begin{center}
\begin{ruledtabular}
\begin{tabular}{l c c}
Motor & \text{Angle $\alpha $} & \text{shape factor  $\mathcal{S}$} \\
\hline
Motor 1 & 43.2$^\circ$ & - 0.618  \\
            & 46.8$^\circ$ &   0.618  \\
\hline            
Motor 2 & 17.7$^\circ$ & 0.465  \\
            & 47.7$^\circ$ & 0.627 \\
\hline            
Motor 3 & 22.4$^\circ$ & 0.638  \\
            & 90.0$^\circ$ & 1.54 
\end{tabular}
\end{ruledtabular}
\caption{For each of the three simple motor protypes, we shape factor $\mathcal{S}$ that corresponds to the optimal settings (shape and location of the rotation axis) is listed. The related shapes are shown in  Fig.\:\ref{fig:motorsshapemax}.
\label{table:shapefactorsimplemotors}}
\end{center}
\end{table}
Considering the constraints put on the shape, Motor 1 and Motor 2 adopt very similar configurations, while the best (convex) shape for Motor 3 is a semi-disk.

%%%%%%%%%%%%%%%%%%%%%%%%%%%%%%%%%%%%%%%%%%%%%%
\subsection{Optimal shape}
%%%%%%%%%%%%%%%%%%%%%%%%%%%%%%%%%%%%%%%%%%%%%%
\label{section:optimalshape}
The three motor realizations show that the angular velocity is sensitive to the precise shape of the motor. We are interested to know what happens if we relax the shape constraints while optimizing for maximum rotation speed. We solve this problem using a numerical procedure.

\begin{figure}
    \begin{center}
        \includegraphics[width=0.9\columnwidth]{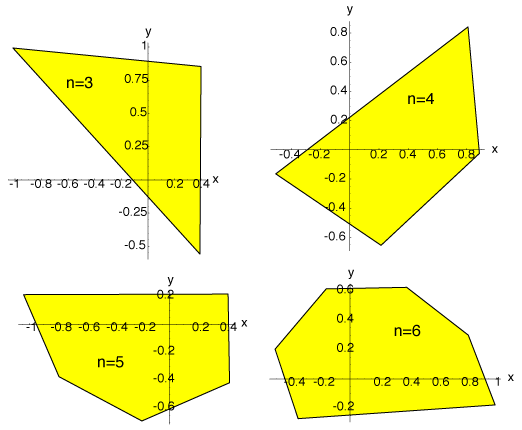}
    \end{center}
    \caption{A numerical procedure was used to find optimum piecewise linear shapes, by determining the best location of each vertex. The results for a small number of vertices, $n = 3, 4, 5, 6$ are shown. Note that the results are invariant for rotations with respect the rotation axis, which is located in the origin $(0, 0)$ of the coordinate system.
        \label{fig:polygonshapecms3456}}
\end{figure}

The boundary of the motor is modeled as piecewise linear. It is defined by the location of $n$ vertices. The numerical procedure finds the optimum location of the $n$ vertices, yielding maximum angular velocity, under the constraints that (1) the mass $M$ remains constant, (2) the shape remains convex. The mass constraint for a homogeneous mass distribution translates into conservation of total area $A$.
For low numbers $n = 3, 4, 5, 6$ the optimum location of the vertices is shown in Fig.\:\ref{fig:polygonshapecms3456}.
Note that the rotation axis is still fixed in the origin $(0,0)$.

\begin{figure}
    \begin{center}
        \includegraphics[width=\columnwidth]{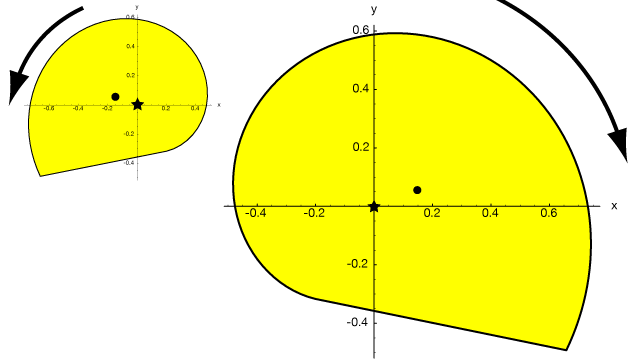}
    \end{center}
    \caption{The (two-dimensional) shape of one motor part that produces the highest average angular velocity, found with a numerical procedure. The rotation axis is marked by a star (at $(0, 0)$), while the center of mass is marked by a dot. The shape in the larger figure rotates from the $y$-axis to the $x$-axis, while its mirror shape (enantiomer) in the smaller figure rotates in the opposite sense. The area of the shape is normalized to 1.
    \label{fig:optimalshape}}
\end{figure}

By increasing the number of vertices $n$, the piecewise linear shape approaches the smooth boundary that yields the highest angular velocity possible.
\begin{figure}
    \begin{center}
        \includegraphics[width=0.9\columnwidth]{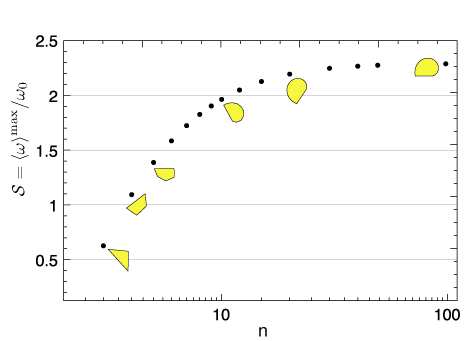}
    \end{center}
    \caption{The shape factor $\mathcal{S}$ is a scale-invariant measure of the influence of the precise geometry of the  motor. For $\mathcal{S} = 0$, the average angular velocity is zero. A numerical procedure was used to obtain the optimum shape (with highest  $\mathcal{S}$) for a motor shape that is piecewise linear, i.\:e.\:consisting of $n$ vertices connected by straight lines. For high $n$ we find an approximation of the exact optimum shape and a lower limit of the maximal $\mathcal{S}$ that can be obtained. $\mathcal{S}$ converges to a value of about 2.29. A negative $\mathcal{S}$ is possible, but this corresponds to a shape that is the mirror image of the shape with opposite $\mathcal{S}$.
         \label{fig:omegamaxn}}
\end{figure}
In Fig.\:\ref{fig:omegamaxn} the shape factor $\mathcal{S}$ is plot against the number of vertices $n$. We see a convergence for large $n$. For $n = 100$ the value of $\mathcal{S}$ is 2.29. This is a factor 3.65 higher than the best value for a triangular shape, $n = 3$, $\mathcal{S} = 0.63$. 
The corresponding shape (for $n = 100$) is shown in Fig.\:\ref{fig:optimalshape}.
A tentative explanation for the optimum spiral shape is that it combines a long curved section with small $r_{\shortparallel}$ (and hence small friction) with a short section that is linear, providing the necessary propulsion.

The chirality of the shape determines the rotation sense. A motor that consists of two identical optimum shapes as shown in the large figure of  Fig.\:\ref{fig:optimalshape} actually has a negative $\mathcal{S}$. This means the motor rotates clockwise (from $y$-axis to $x$-axis). Its enantiomer (the small figure) has positive $\mathcal{S}$ and rotates counterclockwise (from $x$-axis to $y$-axis).

We initially applied the numerical procedure to identical shapes in the two reservoirs. We knew we would simultaneously find the optimum shape for the construction with mirrored shapes in both reservoirs as they share the same shape factor $\mathcal{S}$ (see Eqs.\:\ref{eq:omegaequalparts} and \ref{eq:omegamirrorparts}). Then we extended the numerical procedure so that the shapes in each reservoir could develop independently. For
a small temperature difference ($\Delta T = 1$ K) the shapes become almost exactly each others mirror image (area difference $A_{1}/A_{2} = 1.00045$). Even for a  large temperature difference ($\Delta T = 100$ K) we find mirror shapes with only a small area discrepancy ($A_{1}/A_{2} = 1.017$). The larger shape is in the reservoir with lower temperature. To obtain a positive rotation sense (from $x$-axis to $y$-axis) of the motor for a construction with mirror images, the shape with positive $\mathcal{S}$ needs to be placed in the cold reservoir , while its mirror shape with negative  $\mathcal{S}$ (but equal in absolute value) should be in the warm reservoir.

%%%%%%%%%%%%%%%%%%%%%%%%%%%%%%%%%%%%%%%%%%%%%%
\subsection{Physical estimates for the Brownian rotor}
%%%%%%%%%%%%%%%%%%%%%%%%%%%%%%%%%%%%%%%%%%%%%%
 
We have now collected all the necessary elements to estimate the physical properties of our motor. 
In a real world example of course many of the features discussed in this paper will only be of qualitative use.

We learned that is is advantageous to use a configuration where the shape of the motor in one reservoir is the mirror image of the shape in the other reservoir (section \ref{section: configurations}). We also found the optimal individual shape (section \ref{section:optimalshape}). In the results we present here we assume these optimizations can be approached.

When we use the physical properties of small protein structures in an environment of water molecules (section \ref{globularproteins}) as the separate building blocks of our motors we arrive at an average angular velocity of about 1500 Hz when driven by a temperature gradient of 0.1 K ($T_{1} = 300$ K, $T_{2} = 300.1$ K). This corresponds to about 230 rotations per second.

%%%%%%%%%%%%%%%%%%%%%%%%%%%%%%%%%%%%%%%%%%%%%%
\section{Cross processes}
%%%%%%%%%%%%%%%%%%%%%%%%%%%%%%%%%%%%%%%%%%%%%%
We discussed a Brownian motor, and derived a relationship between its motion -- the average angular velocity $\langle \omega \rangle$ -- and the applied temperature difference $\Delta T$. This relationship is an example of a \emph{cross} process.
The \emph{normal} process that would give rise to a motion $\langle \omega \rangle$ originates from a mechanical force. In our system with only a rotational degree of freedom this mechanical force would be in the form of a torque $\tau$ along the $z$-axis.

Cross processes are very common in physics. One well-known example is the Seebeck effect \cite{callen}, where a temperature difference over an electric conductor causes an electric current. The Seebeck effect has a reverse or \emph{mirror} cross process: the Peltier effect. Here, an applied electric current causes a temperature difference. Processes and their mirror processes are  related through a general principle of stability. In the example of the Seebeck-Peltier effects, consider a system that is originally in equilibrium. If it is perturbed by the application of a temperature gradient, currents will start to flow (the Seebeck effect), which in turn will give rise to a counteracting temperature difference (the Peltier effect), attempting to cancel out the original cause of the disturbance. Another example is Lenz' law in electromagnetism. Moving a ferromagnetic core into a coil will induce currents in the coil. These currents are such that the resulting magnetic field will expel the core, hence counteracting the original disturbance.

Now that we established the existence of a cross process in the Brownian motor system and showed the relationship between a cross process and its mirror process, the question naturally arises: what is the mirror process in our system? It should be one that counteracts the original perturbation -- a temperature difference between the reservoirs. A flow of heat $\dot{Q}_{1  \rightarrow 2}$ from reservoir 1 to reservoir 2  would do exactly that. And we know the \emph{normal} process (that causes rotational motion) is induced by a torque $\tau$. The \emph{mirror} cross process then is a heat flow caused by a torque. When we perturb our system in temperature equilibrium by applying a torque $\tau$, the motor will of course rotate. How can the system react to counterbalance this motion? By creating a temperature gradient between the two reservoirs, that according to the theory of the Brownian motor will cause rotational motion. The sign of the temperature difference is such that the resulting rectified Brownian motion opposes the motion started by the torque.

%%%%%%%%%%%%%%%%%%%%%%%%%%%%%%%%%%%%%%%%%%%%%%
\section{Model of the Brownian Refrigerator}
%%%%%%%%%%%%%%%%%%%%%%%%%%%%%%%%%%%%%%%%%%%%%%
\begin{figure}
    \begin{center}
        \includegraphics[width=0.8\columnwidth]{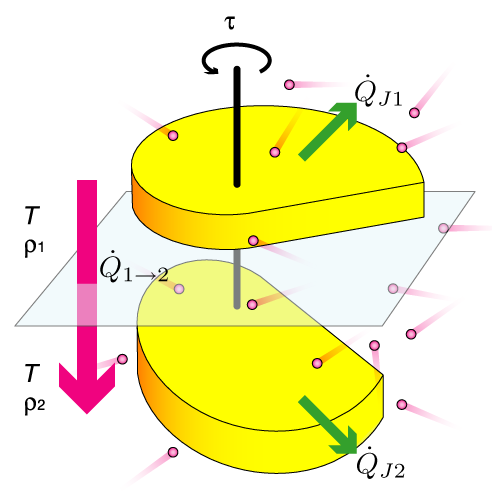}
    \end{center}
    \caption{The Brownian refrigerator is a rotating device consisting of two connected bodies, each in a temperature reservoir. By applying a torque on the apparatus, a heat flow $\dot{Q}_{1  \rightarrow 2}$ will arise that can cool down reservoir 1 at the expense of heating up reservoir 2. It is possible to obtain the conditions where this cooling power is larger than the dissipative heating $\dot{Q}_{J 1}$.
\label{fig:modelrefrigerator}}
\end{figure}
We propose a cooling apparatus or refrigerator based on the Brownian motor described before. 
Two reservoirs (see Fig.\:\ref{fig:modelrefrigerator}) are thermally isolated from each other and initially at temperature equilibrium, $T_{1} = T_{2} = T$.
The refrigerator consists of a rotating device with a part in each reservoir. The parts are rigidly connected via a rotation axis and are subject to random collisions with particles in their reservoirs. These have mass $m$, considered much smaller than the total mass $M$ of the rotating device.
On applying a torque $\tau$ along the $z$-axis the system will develop a heat flow  $\dot{Q}_{1  \rightarrow 2}$, cooling reservoir 1 down at the expense of heating up reservoir 2. 
The following derivation of this heat flow is done for a construction where one part of the device is the reflected copy of the other (as in Fig.\:\ref{fig:3configurations}b) because it shows a linear response in small temperature differences $T_{1} - T_{2}$ and it produces the highest rotating speeds and resulting heat flow.

%%%%%%%%%%%%%%%%%%%%%%%%%%%%%%%%%%%%%%%%%%%%%%
\section{Linear response and Onsager symmetry}
%%%%%%%%%%%%%%%%%%%%%%%%%%%%%%%%%%%%%%%%%%%%%%
Previously we derived a relation between the average angular velocity $\langle \omega \rangle$ of the rotating brownian motor and the temperatures of the two isolated reservoirs, $T_{1}$ and $T_{2}$, correct to order $m/M$:
\begin{equation}
\langle \omega \rangle
= \frac{\sqrt{2 \pi k_{B} m}}{4I}
\frac{\rho_{1} \rho_{2} (T_{2}^{1/2} + T_{1}^{1/2})(T_{2} - T_{1})}
{(\rho_{1} T_{1}^{1/2} + \rho_{2} T_{2}^{1/2})^{2}}
\frac{\oint dl\,  r_{\shortparallel}^{3}}{\oint  dl\,  r_{\shortparallel}^{2}}.
\end{equation}
Here $I$ is the inertial moment of the motor with respect to its rotation axis and $\oint dl\,  r_{\shortparallel}^{3}/\oint  dl\,  r_{\shortparallel}^{2}$ is a geometrical factor defined by the shape of the rotating motor parts. The geometrical factor is zero in a symmetrical configuration, hence the importance of choosing appropriate asymmetric or chiral elements. For a more detailed discussion of the geometrical factor we refer to Section \ref{section:shape}.
For a small temperature difference $\Delta T$ between the two reservoirs,
\begin{alignat}{2}
T_{1} &= T - \Delta T / 2, &\quad \Delta T &= T_{2} - T_{1} \ll T,\\
T_{2} &= T + \Delta T / 2, &\quad T &= (T_{1} + T_{2})/2,
\end{alignat}
the mechanical response $\langle \omega \rangle$ is linear in $\Delta T$ to very good approximation:
\begin{equation}
\langle \omega \rangle
\approx \frac{\sqrt{2 \pi k_{B} m}}{2I}
\frac{\rho_{1} \rho_{2}}
{(\rho_{1} + \rho_{2})^{2}}
\frac{\oint dl\,  r_{\shortparallel}^{3}}{\oint  dl\,  r_{\shortparallel}^{2}}
\frac{\Delta T}{T^{1/2}}.
\label{linearresponse}
\end{equation}

An elegant way to calculate the cooling potential of our system is by making use of Onsager's relations \cite{onsager}. 
We will identify a \emph{flow} and a \emph{force} for the two cross processes involved.The alternative is to revisit the analysis of Section \ref{fluctutions}, adding an extra torque term to the master equation. For the \emph{mechanical} process we identify a flow $J_{1} = \langle \omega \rangle$ and a thermodynamic force $X_{2} = \Delta T / T^{2}$  in the linear relation of Eq.\:\ref{linearresponse}.
The proportionality constant $L_{12}$ of the first Onsager relation,
\begin{equation}
J_{1} = L_{12} X_{2},
\end{equation}
is, for our particular system, given by
\begin{equation}
L_{12}
= \frac{\sqrt{2 \pi k_{B} m}}{2I}
\frac{\rho_{1} \rho_{2}}{(\rho_{1} + \rho_{2})^{2}} T^{3/2}
\frac{\oint dl\,  r_{\shortparallel}^{3}}{\oint  dl\,  r_{\shortparallel}^{2}}.
\end{equation}

For the second cross process, the \emph{cooling} process, we can identify a heat flow $J_{2} = \dot{Q}_{1  \rightarrow 2}$.
The force $X_{1}$ is given by the chemical potential associated with the particle flow $J_{1} = \langle \omega \rangle$ of the normal process, and is generated by the application of the torque $\tau$. More precisely, $X_{1} = \tau / T$. Again we expect a linear response
\begin{equation}
J_{2} = L_{21} X_{1}.
\end{equation}

Onsager symmetry now tells us that the two proportionality coefficients of the cross processes are identical,
\begin{equation}
L_{21} = L_{12}.
\end{equation}
The heat flow from one reservoir to the other in the linear regime now becomes obvious,
\begin{equation}
\dot{Q}_{1  \rightarrow 2}
= \frac{\sqrt{2 \pi k_{B} m T}}{2I}
\frac{\rho_{1} \rho_{2}}{(\rho_{1} + \rho_{2})^{2}}
\frac{\oint dl\,  r_{\shortparallel}^{3}}{\oint  dl\,  r_{\shortparallel}^{2}}
\, \tau.
\label{heattorque}
\end{equation}

The complete Onsager relations, combining normal and cross processes, are given by:
\begin{equation}
J_{1} = L_{11} X_{1} +  L_{12} X_{2}, \quad 
J_{2} = L_{21} X_{1} +  L_{22} X_{2}.
\end{equation}
The Onsager coefficients we have not identified yet are:
\begin{equation}
L_{11} = T / \gamma, \quad 
L_{22} = \frac{\gamma_{1} \gamma_{2} k_{B} T^{2}}{\gamma I},
\end{equation}
while $J_{1} =  \langle \omega \rangle$, $J_{2} = \dot{Q}_{1  \rightarrow 2}$, $X_{1} = \tau / T$, and $X_{2} = \Delta T / T^{2}$ as before. $L_{11}$ can be associated to the direct mechanical response of the motor to the application of a torque, while $L_{22}$ is related to heat conductivity, being the coefficient between the heat flow and the temperature gradient.

%%%%%%%%%%%%%%%%%%%%%%%%%%%%%%%%%%%%%%%%%%%%%%
\section{Results and discussion}
%%%%%%%%%%%%%%%%%%%%%%%%%%%%%%%%%%%%%%%%%%%%%%
 \begin{figure}
    \begin{center}
        \includegraphics[width=0.9\columnwidth]{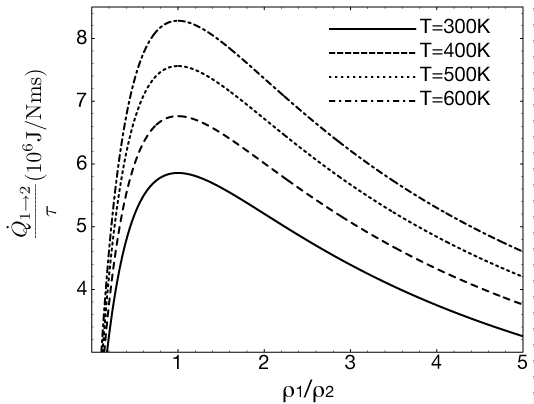}
    \end{center}
    \caption{Physical values of the cooling power $\dot{Q}_{1  \rightarrow 2}$ of the Brownian refrigerator in response to a torque $\tau$ are given for an environment of two nitrogen gas reservoirs for several choices of the temperature $T$ of the gases. The device works best when the densities of the gases, $\rho_{1}$ and $\rho_{2}$, are equal. The mass of the gas particles is $m = 5\times 10^{-26} \text{kg}$, while the mass of the device is $M = 1.66\times10^{-22}  \text{kg}$.
    \label{fig:QoverTau}}
\end{figure}
Eq.\:\ref{heattorque} gives us a relation between the heat flow from reservoir 1 to reservoir 2 and the applied torque.
Earlier we chose the parameters of the building blocks that constitute our motor and its environment to be comparable with globular proteins in water, see Table\:\ref{table:globularprotein} for details.
Using these variables again for the refrigerating device we obtain a heat flow $\dot{Q}_{1  \rightarrow 2}$ of 4.5 $\mu$J/s for every pNm of torque applied. This result is for $T = 300$ K and a shape that is close to optimal. Note that in the ratio $\dot{Q}_{1  \rightarrow 2} / \tau$, the geometrical factor,
\begin{equation}
\frac{M}{I}
\frac{\oint dl\,  r_{\shortparallel}^{3}}{\oint  dl\,  r_{\shortparallel}^{2}},
\end{equation}
found earlier reappears so that the comprehensive discussion therein (sections \ref{section:shape} and \ref{section:optimalshape}) remains applicable for the various shapes the parts of the refrigerator can adopt.

Fig.\:\ref{fig:QoverTau} shows the dependency of the cooling power $\dot{Q}_{1  \rightarrow 2} / \tau$ on the densities $\rho_{1}$ and $\rho_{2}$ of the gas reservoirs for different gas temperatures $T$. Here we have set the cooling device in the membrane separating two gaseous environments (the values for nitrogen gas $\text{N}_{2}$ are  used: $m = 5\times 10^{-26} \text{kg}$, while for the refrigerator $M = 1.66\times10^{-22}  \text{kg}$ and typical radius 3 nm). The size of the effect is determined by the ratio of the gas densities $\rho_{1}/\rho_{2}$. For all temperatures maximal cooling power is found when the densities in the two reservoirs are equal, $\rho_{1} = \rho_{2}$. Higher heat fluxes arise when the gas temperature $T$ is higher.

The maximal torque that can be applied and the maximum obtainable temperature difference will be determined by increasing dissipative heat flows, which we discuss now.

%%%%%%%%%%%%%%%%%%%%%%%%%%%%%%%%%%%%%%%%%%%%%%
\subsection{Joule dissipation}
%%%%%%%%%%%%%%%%%%%%%%%%%%%%%%%%%%%%%%%%%%%%%%

We showed the occurrence of a heat flux $\dot{Q}_{1  \rightarrow 2}$ that takes away heat from reservoir 1 to reservoir 2.
Friction however will cause Joule heating in both reservoirs $i = 1,2$ by an amount
\begin{equation}
\dot{Q}_{J i}
= \gamma_{i} \langle \omega \rangle^{2}
= \gamma_{i} \tau^{2} / \gamma^{2}.
\label{jouleheating}
\end{equation}
If we want reservoir 1 to cool down, the heat $\dot{Q}_{1  \rightarrow 2}$ transferred from reservoir 1 to 2 needs to be larger than the heat $\dot{Q}_{J 1}$ dissipated by friction in reservoir 1,
\begin{equation}
\frac{\dot{Q}_{1  \rightarrow 2}}{\dot{Q}_{J 1}}
= 2 \frac{k_{B} T m}{I} \frac{\rho_{2}}{\tau} \oint dl\,  r_{\shortparallel}^{3} > 1.
\end{equation}
This condition poses a limit on the applied torque:
\begin{equation}
\tau
< \tau_{\text{lim}} =
 2 \frac{k_{B} T m}{I} \rho_{2} \oint dl\,  r_{\shortparallel}^{3}.
\end{equation}
A greater torque would cause the heat dissipation to annihilate the cooling effect.
We are able to suggest a scale-invariant numerical limit for the torque. Note that both the inertial moment $I$ and the shape factor $\oint dl\,  r_{\shortparallel}^{3}$ scale with $R^{4}$ ($R$ being the typical linear dimension of the refrigerator) in the case that the refrigerator parts have homogeneous density $\rho_{m}$. If we assume optimum operation using the appropriate  chiral shapes of the parts, as will be developed in detail later (section\:\ref{section:maximumcooling}), we find
\begin{equation}
\rho_{m} \frac{\oint dl\,  r_{\shortparallel}^{3}}{I}   = 1.30684.
\end{equation}
This result depends only on the geometry and not on the dimensions of the parts. It will be lower for less favorable shapes and zero for a symmetric construction.
The maximal torque than can be expressed as
\begin{equation}
\tau_{\text{lim}}
= 2 k_{B} T  \frac{\rho_{2} m}{\rho_{m}} \times 1.30684.
\end{equation}
This maximal torque is proportional to the ratio of the mass density $\rho_{2} m$ of gas that is heated and  of the refrigerator $\rho_{m}$. It does not depend on the size of cooling device.
For $T = 300$ K and a system according to Table\:\ref{table:globularprotein}, assuming optimum shape, 
\begin{equation}
\tau_{\text{lim}}
= 3.92\times 10^{-21} \text{Nm}.
\end{equation}

%%%%%%%%%%%%%%%%%%%%%%%%%%%%%%%%%%%%%%%%%%%%%%
\subsection{Maximal net cooling}
%%%%%%%%%%%%%%%%%%%%%%%%%%%%%%%%%%%%%%%%%%%%%%
\label{section:maximumcooling}
The cooling power of the refrigerator is proportional to the applied torque $\tau$ (Eq.\:\ref{heattorque}), while the dissipative heat flux grows with $\tau^{2}$ (Eq.\:\ref{jouleheating}).
For large $\tau$ the cooling effect will be annihilated by dissipation and in the previous section we calculated a cut-off $\tau_{\text{lim}}$, at which both effects cancel each other.
We can also calculate the torque $\tau_{\text{max}}$ that maximizes the net cooling,
\begin{equation}
\dot{Q}_{\text{net}} = \dot{Q}_{1  \rightarrow 2} - \dot{Q}_{J 1}.
\label{netcooling}
\end{equation}

A simple calculation leads to a maximum of $\dot{Q}_{\text{net}} = A \tau - B \tau^{2}$ at $\tau_{\text{max}} = \tau_{\text{lim}} / 2 = A/(2B)$, with $A = (\sqrt{2 \pi k_{B} m T})/(2I)
(\rho_{1} \rho_{2})/((\rho_{1} + \rho_{2})^{2})
(\oint dl\,  r_{\shortparallel}^{3}) / (\oint  dl\,  r_{\shortparallel}^{2})$ and $B = \gamma_{1} / \gamma$.
For the optimum torque then we find:
\begin{equation}
\tau_{\text{max}}
= \frac{k_{B} T m}{I} \rho_{2} \oint dl\,  r_{\shortparallel}^{3}.
\end{equation}
This result, like $\tau_{\text{lim}}$, is independent of the size of the refrigerator (assuming a homogeneous interior). It only depends on the density of the the environment and the refrigerator, the environment temperature and the specific shape of the refrigerator. For an optimum shape (see later) and variables according to Table\:\ref{table:globularprotein} we find:
\begin{equation}
\tau_{\text{max}}
= 1.96\times 10^{-21} \text{Nm}.
\end{equation}

Substituting the explicit expression for $\tau_{\text{max}}$ into Eq.\:\ref{netcooling} yields the maximal net heat flow out of reservoir 1:
\begin{equation}
\dot{Q}_{\text{net}}^{\text{max}}
= \frac{\sqrt{\pi}}{8} \frac{(k_{B} T m)^{3/2}}{I^{2}}
\frac{\rho_{1} \rho_{2}^{2}}{(\rho_{1} + \rho_{2})^{2}}
\frac{\left(\oint dl\,  r_{\shortparallel}^{3}\right)^{2}}{\oint  dl\,  r_{\shortparallel}^{2}}.
\label{Qnetmax}
\end{equation}
We see that when we also take into account the loss through friction, the refrigerator is most effective when the densities are the same in both reservoirs, $\rho_{1} = \rho_{2}$. The net cooling is higher when the device works in higher gas densities. The reason may seem counterintuitive: the heat dissipation through friction is smaller in an environment with higher friction. Eq.\:\ref{jouleheating} shows that the dissipated heat is proportional to the square of the average angular velocity $\langle \omega \rangle$ that is obtained by applying a torque $\tau$, and this velocity is lower when the friction $\gamma$ is higher.

\begin{figure}
    \begin{center}
        \includegraphics[width=0.8\columnwidth]{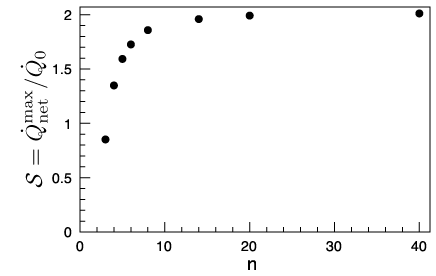}
    \end{center}
    \caption{The effectiveness of the refrigerator is influenced by its (chiral) shape. $\mathcal{S}$ gives a size-independent measure of this geometrical factor. With a numerical procedure we can optimize this factor, by approximating the shape of the motor parts as piecewise linear, with $n$ edges. For $n$ sufficiently high we approach the highest obtainable factor $\mathcal{S}$, and the corresponding shape is represented in Fig.\:\ref{fig:refrigeratorshape}.
    \label{fig:nshape}}
\end{figure}

\begin{figure}
    \begin{center}
        \includegraphics[width=0.8\columnwidth]{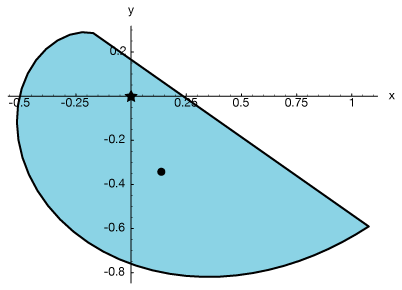}
    \end{center}
    \caption{The shape of one part of the refrigerating device that maximizes the net cooling power, which also takes into account the frictional dissipation. The complete device consists of this shape in one reservoir and the \emph{mirror} image of this shape in the other. The two parts are rigidly connected with each other by a rotation axis, marked by a star. The mass distribution in the interior of the parts is assumed homogeneous. The center of mass is represented by a dot.
    \label{fig:refrigeratorshape}}
\end{figure}

Before we can give numerical results, we need to investigate the role of the geometry of the refrigerator. We separate a scale-invariant shape factor from Eq.\:\ref{Qnetmax},
\begin{equation}
\mathcal{S} =
\frac{1}{\sqrt{A}} \frac{M^{2}}{I^{2}}
\frac{\left(\oint dl\,  r_{\shortparallel}^{3}\right)^{2}}{\oint  dl\,  r_{\shortparallel}^{2}},
\label{shapefactor2d}
\end{equation}
where $A$ is the area of one part of the construction.
All the factors that depend on the shape, the contours and the inertial moment, are included in $\mathcal{S}$.
We will begin straight away with optimizing this shape factor, and for most of the realizations of the refrigerator presented in this paper we assume the shape is (close to) optimum. 
The calculations however can also be done for other (less favorable) shapes; the elements of the calculation are comparable to those presented earlier, where we also discussed three simple model shapes that can be analyzed analytically.
A similar numerical procedure as was used to find the optimal shape of the motor (see section\:\ref{section:optimalshape}), yields a value for $\mathcal{S}$, see Fig.\:\ref{fig:nshape}.
The procedure approximates the shape of the refrigerator parts as piecewise linear with $n$ vertices.
For sufficiently large $n$, $\mathcal{S}$ converges to a value slightly higher than 2.
The corresponding optimum shape for the cooling function is then also found, see Fig.\:\ref{fig:refrigeratorshape}.
The shape is that of a part of the refrigerator in one reservoir; placing the mirrored shape in the other reservoir gives the optimum configuration of the refrigerator. The axis of rotation is given by the $z$-axis.
The numerical procedure expects the mass $M$ of the motor to be homogeneously distributed, which reflects the reality of biological entities such as proteins that could function as parts of the device.

For a homogeneous mass distribution we write the mass of the refrigerator as $M = 2 \rho_{m} A$, with $\rho_{m}$ the constant density of the refrigerator parts.
To find the maximum net heat flux the numerical factor $\mathcal{S}$ is multiplied by
\begin{equation}
\dot{Q}_{0}
= \frac{\sqrt{2 \pi}}{16} \frac{(k_{B} T m)^{3/2}}{\rho_{m}^{2} A^{3/2}}
\frac{\rho_{1} \rho_{2}^{2}}{(\rho_{1} + \rho_{2})^{2}},
\end{equation}
so that $\dot{Q}_{\text{net}}^{\text{max}} = \dot{Q}_{0} \mathcal{S}$.
Remember that our theory is two-dimensional. The result however depends on the absolute values of the gas densities $\rho_{1}$ and $\rho_{2}$, contrary to Eq.\:\ref{heattorque} where only their relative magnitudes play a part . We cannot simply insert values for three-dimensional gas densities. Therefore we make a small technical detour to describe the three-dimensional expressions and justify for which case the two-dimensional shape optimization remains valid.

%%%%%%%%%%%%%%%%%%%%%%%%%%%%%%%%%%%%%%%%%%%%%%
\subsection{Three-dimensional model}
%%%%%%%%%%%%%%%%%%%%%%%%%%%%%%%%%%%%%%%%%%%%%%
In a three-dimensional analysis the essential difference is the description of the geometrical factors. Instead of a contour integral $\oint dl$ there is an integral $\int dS$ over the surface of the refrigerator parts, while the vector $r_{\shortparallel}$ gets a new definition, see Section\:\ref{section:3dmotor} for details.
For the ratio of the cooling power over the applied torque we find
\begin{equation}
\frac{\dot{Q}_{1  \rightarrow 2}}{\tau}
= \frac{\sqrt{2 \pi k_{B} m T}}{2I}
\frac{\rho_{1} \rho_{2}}{(\rho_{1} + \rho_{2})^{2}}
\frac{\int dS\,  r_{\shortparallel}^{3}}{\int  dS\,  r_{\shortparallel}^{2}},
\end{equation}
while the maximum net cooling power of reservoir 1 now becomes
\begin{equation}
\dot{Q}_{\text{net}}^{\text{max}}
= \frac{\sqrt{\pi}}{8} \frac{(k_{B} T m)^{3/2}}{I^{2}}
\frac{\rho_{1} \rho_{2}^{2}}{(\rho_{1} + \rho_{2})^{2}}
\frac{\left(\int dS\,  r_{\shortparallel}^{3}\right)^{2}}{\int  dS\,  r_{\shortparallel}^{2}}.
\end{equation}
We again use a product of an external $\dot{Q}_{0}$ and a scale-invariant geometrical factor $\mathcal{S}$,
\begin{equation}
\dot{Q}_{\text{net}}^{\text{max}} = \dot{Q}_{0} \mathcal{S}.
\end{equation}
The geometrical factor now needs to be scaled by a factor proportional to $R^{2}$, with $R$ the typical linear dimension of the shape. For this we use $V^{2/3}$, with $V$ the volume of a refrigerator part:
\begin{equation}
\mathcal{S} =
\frac{1}{V^{2/3}} \frac{M^{2}}{I^{2}}
\frac{\left(\int dS\,  r_{\shortparallel}^{3}\right)^{2}}{\int  dS\,  r_{\shortparallel}^{2}},
\label{shapefactor3d}
\end{equation}
The shape factor can then be solved analytically or numerically, producing size-independent results.
The remaining factor
\begin{align}
\dot{Q}_{0}
&= \frac{\sqrt{\pi}}{8} \frac{(k_{B} T m)^{3/2}}{M^{2}}
\frac{\rho_{1} \rho_{2}^{2}}{(\rho_{1} + \rho_{2})^{2}}
V^{2/3}
\\
&= \frac{\sqrt{2 \pi}}{16} \frac{(k_{B} T m)^{3/2}}{\rho_{m}^{2} V^{4/3}}
\frac{\rho_{1} \rho_{2}^{2}}{(\rho_{1} + \rho_{2})^{2}},
\label{Q03d}
\end{align}
then shows again the role of the various parameters. In the second expression for $\dot{Q}_{0}$ we substituted $M = 2 \rho_{m} V$ for the homogeneous case. 

We now argue that we can recuperate the two-dimensional optimization of the shape factor $\mathcal{S}$ and corresponding two-dimensional shape. We propose a prismatic structure for each of the refrigerator parts, defined by two equal, flat (two-dimensional) surfaces separated by a distance $H$, as shown in the initial model of the device in Fig.\:\ref{fig:modelrefrigerator}. The rotation axis is perpendicular to the two surfaces. For this configuration it is easy to show that the surface integrals can be written as the product of the distance $H$ and the contour integral of the top or bottom surface, exactly as in the two-dimensional description:
\begin{align}
\int  dS\,  r_{\shortparallel}^{2}
&= H \oint dl\,  r_{\shortparallel,2D}^{2},
\\
\int  dS\,  r_{\shortparallel}^{3}
&= H \oint dl\,  r_{\shortparallel,2D}^{3}.
\end{align}
Also, for a homogeneous mass distribution,
\begin{equation}
\frac{M}{I} = \frac{V}{\int r_{\perp}^{2} dV} 
= \frac{H A}{H \int r^{2} dA},
\end{equation}
where we take the integral over the volume $V$ of $r_{\perp}$, the distance to the rotation axis, and find it is equivalent to taking the surface integral over the top or bottom surface $A$ of the prism times the thickness $H$.
In conclusion we recover the two-dimensional shapefactor (Eq.\:\ref{shapefactor2d}) by inserting the corresponding volume $V = (H \sqrt{A})^{3/2}$ in Eq.\:\ref{shapefactor3d}. The approach is then to use the three-dimensional expression for $\dot{Q}_{0}$  (Eq.\:\ref{Q03d}), and the numerical results for the two-dimensional case, as in Fig.\:\ref{fig:nshape}. 

 \begin{figure}
    \begin{center}
        \includegraphics[width=0.9\columnwidth]{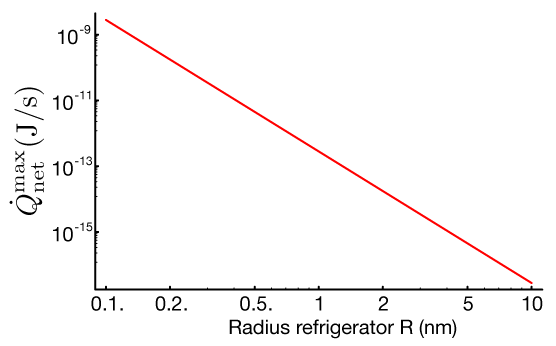}
    \end{center}
    \caption{The net cooling power of the Brownian refrigerator (the dissipative loss by friction is deducted from the heat flux that cools down one compartment) depends strongly on the size of the cooling device.
Values for a device operating between two aquatic reservoirs ($m = 2.992\times10^{-26}  \text{kg}$, $\rho_{1} = \rho_{2} = 3.343\times 10^{28} \text{m}^{-3}$) at temperature $T = 300 \text{K}$ are presented here. The size-dependency of the mass of the rotating device $M$ is included, while its density is that of a typical protein, $\rho_{m} = 1380 \:\text{kg m}^{-3}$.
          \label{fig:QR3D}}
\end{figure}

Doing this for a temperature ($T = 300 \text{K}$), reservoir densities ($\rho_{1} = \rho_{2} = 3.343\times 10^{28} \text{m}^{-3}$) and  $m = 2.992\times10^{-26}  \text{kg}$ for an aquatic environment, and $\rho_{m} = 1380 \:\text{kg m}^{-3}$, typical for proteins, $\dot{Q}_{\text{net}}^{\text{max}} = \dot{Q}_{0} \mathcal{S}$ can be expressed as a function of only the radius $R$ of one refrigerator part, as in Fig.\:\ref{fig:QR3D}. 
For a globular protein of typical dimension $R \approx 3 \text{nm}$, we find a value of $\dot{Q}_{\text{net}}^{\text{max}} = 3.5 \times 10^{-15} \text{J}/\text{s}$. Note the strong size dependence: a refrigerator of one nm, would yield a cooling power of about  $2.8 \times 10^{-13} \text{J}/\text{s}$. For comparison, it takes about $2.2 \times 10^{-12} \text{J}$ to cool down a 1 $\mu\text{m}$ small cell one Kelvin, which could be accomplished by one refrigerator of 1 nm radius in one minute.

%%%%%%%%%%%%%%%%%%%%%%%%%%%%%%%%%%%%%%%%%%%%%%
\subsection{Thermal conductivity}
%%%%%%%%%%%%%%%%%%%%%%%%%%%%%%%%%%%%%%%%%%%%%%
As mentioned earlier, the $L_{22}$ Onsager coefficient can be related to heat conductivity between the reservoirs.
In the linear response model, the heat conducted from one reservoir to the other can be quantified:
\begin{equation}
\dot{Q}_{\text{cond}} = \frac{\gamma_{1} \gamma_{2}}{\gamma I} k_{B} \Delta T.
\end{equation}
Here $\gamma_{1}$ and $\gamma_{2}$ are the friction coefficients of the separate parts of the refrigerator, and $\gamma = \gamma_{1} + \gamma_{2}$ represents the total friction coefficient.
\begin{figure}[h]
    \begin{center}
        \includegraphics[width=0.8\columnwidth]{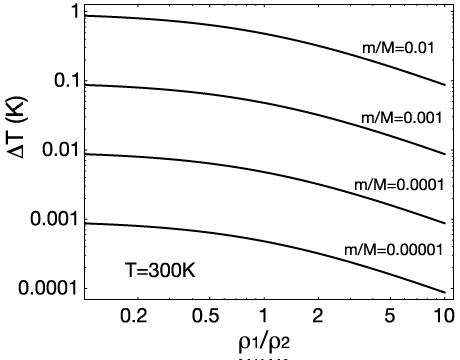}
    \end{center}
    \caption{The temperature gradient $\Delta T$ for which the conductive heat flow cancels the cooling heat flow. For the device to cool down reservoir 1 (at the expense of reservoir 2), the temperature difference must remain under this limit. The different curves are for examples of the mass ratios between colliding particle ($m$) and device $M$. A small $M$ (or large $m$) is beneficial. The sustainable temperature gradient is highest for high gas density $\rho_{2}$ in reservoir 2  (relative to $\rho_{1}$).  For high gas density $\rho_{1}$ in reservoir 1 the obtainable temperature gradient becomes very small. The graph is for absolute temperature differences around $T = 300$ K.
    \label{fig:thermalconductivity}}
\end{figure}
The conductive heat flow is proportional to the temperature gradient $\Delta T$ and goes from the warm to the cold reservoir. Therefore it is directed against the cooling power of the device and for a temperature difference larger than $\Delta T_{\text{lim}}$ the net cooling effect will vanish. The condition $\dot{Q}_{\text{cond}} < \dot{Q}_{\text{net}}^{\text{max}}$ leads to
\begin{equation}
\frac{\Delta T}{T} <
\frac{\pi}{8}\frac{m}{I} \frac{\rho_{2}}{\rho_{1} + \rho_{2}}
\frac{\left(\oint dl\,  r_{\shortparallel}^{3}\right)^{2}}
{\left(\oint dl\,  r_{\shortparallel}^{2}\right)^{2}}.
\end{equation}
Note that the sustainable temperature gradient is proportional to $m/M$. In Fig.\:\ref{fig:thermalconductivity} we show $\Delta T_{\text{lim}}$ for varying $m/M$ as a function of the density ratio $\rho_{1} / \rho_{2}$. For our choice of the direction of the cooling effect (reservoir 1 cools down, reservoir 2 heats up), a higher relative density in reservoir 2  allows a larger $\Delta T_{\text{lim}}$. In the limit of $\rho_{1}$ being much smaller than $\rho_{2}$, $\Delta T_{\text{lim}} / T$ approaches 
$(\pi / 8) (m / I)
\left(\oint dl\,  r_{\shortparallel}^{3}\right)^{2} / 
\left(\oint dl\,  r_{\shortparallel}^{2}\right)^{2}$
Earlier we found  equal densities, $\rho_{1} = \rho_{2}$,  to correspond with maximum net cooling power, for which the maximum sustainable temperature gradient $\Delta T_{\text{lim}}$ is half this value. 

For the previously used example of globular proteins (see Table\:\ref{table:globularprotein} for the parameters), $\Delta T_{\text{lim}} = 4.3$ mK at $T = 300$ K and equal reservoir densities.

In all examples given, the shape is presumed to be (close to) optimal.

%%%%%%%%%%%%%%%%%%%%%%%%%%%%%%%%%%%%%%%%%%%%%%
\appendix
%%%%%%%%%%%%%%%%%%%%%%%%%%%%%%%%%%%%%%%%%%%%%%

%%%%%%%%%%%%%%%%%%%%%%%%%%%%%%%%%%%%%%%%%%%%%%
\section{Full expressions for model motors}
%%%%%%%%%%%%%%%%%%%%%%%%%%%%%%%%%%%%%%%%%%%%%%
For Motor 1 a straightforward calculation leads to
\begin{align}
\oint  dl\,  r_{\shortparallel}^{2} &=
\frac{R}{72} \biggl(R^2 \cos 4 \alpha 
\nonumber\\
&+ 2 R \bigl( (R+6 y) \cos 3 \alpha - (R+6 x) \sin 3\alpha \bigr)
\nonumber\\
&+ 12 \bigl( (y-x) (R+3 (x+y)) \cos 2 \alpha  - 6 x y \sin 2 \alpha \bigr)
\nonumber\\
&+ 6 \bigl( (R^2+2 y R + 12 y^2) \cos \alpha 
\nonumber\\
&+  (R^2+2 x R+12 x^2) \sin \alpha \bigr)
\nonumber\\
&+ 7 R^2+12 (x+y) R + 36 (x^2+y^2)\biggr),
\label{eq:motor1int2}
\end{align}
while $I / M = R^{2}/18 + x^{2} + y^{2}$.
The friction coefficient is then directly obtained through Eq.~(\ref{eq:frictioncoefficient}). One can note that this coefficient is always nonzero, unless $R = 0$. Also for $\alpha = 0$, when the motor has the shape of two connected bars, Eq.~(\ref{eq:motor1int2}) is nonzero. Choosing $\alpha = 0$, $x = 0$, and $y = R / 3$, we retrieve the expression for a rod of length $R$ rotating about one end,
\begin{equation}
\oint  dl\,  r_{\shortparallel}^{2} = 2 R^{3} / 3.
\end{equation}
The sense of rotation is geometrically determined by
\begin{align}
\oint  dl\,  r_{\shortparallel}^{3} &=
-\frac{R \sin 2 \alpha}{432} \biggl( R^3 \sin 4 \alpha
\nonumber\\
&+ 18 R^2 \bigl( x \cos 3 \alpha  +  y \sin 3 \alpha \bigr)
\nonumber\\
&+ 108 R \bigl( 2 x y \cos 2 \alpha  - (x^{2}-y^{2}) \sin 2 \alpha  \bigr)
\nonumber\\
&+ 54 \bigl( x (R^2-4 x^2+12 y^2) \cos \alpha
\nonumber\\
&- y (R^2+12 x^2-4 y^2) \sin \alpha  \bigr) \biggr).
\label{eq:motor1int3}
\end{align}
For a bar-shaped motor, $\alpha = 0$ or $\alpha = 90^\circ$, Eq.~(\ref{eq:motor1int3}) effectively returns zero, resulting is a zero average rotation for the motor. The shape needs to have chiral asymmetry for the motor to function. This observation becomes more apparent when we locate the rotation axis in the center of mass, $(x,y) =(0,0)$. Then Eq.~(\ref{eq:motor1int3}) simply becomes
\begin{equation}
\oint  dl\,  r_{\shortparallel}^{3} = -\frac{R^4}{432} \sin (2 \alpha ) \sin (4 \alpha ),
\end{equation}
which equals zero also when $\alpha = 45^\circ$. For this angle the shape contains an extra symmetry axis, lifting the chiral symmetry. To find a nonzero rotation, we need to change the shape to a chiral asymmetric one. We see an opposite rotation sense when moving across the symmetric position.  In particular for this choice of the position of the rotation axis, we note that $\langle \omega \rangle[\pi/2 - \alpha] = - \langle \omega \rangle[\alpha]$, where for this triangular shape, $\alpha$ and $\pi/2 - \alpha$ correspond to chiral enantiomers. This is because $\oint  dl\,  r_{\shortparallel}^{2}$, $M / I$, and $v_{0}$ are even under a $\alpha \rightarrow  \pi/2 - \alpha$ transformation, while $\oint  dl\,  r_{\shortparallel}^{3}$ is uneven.
For other locations of the rotation axis, we see a similar behavior. In Fig.÷ \ref{fig:contoursmotor1} we show the dependency of the angular velocity $\langle \omega \rangle$ on the exact place of the axis (which in our calculations can be outside the motor body). For chiral shapes, the axis of rotation that yields zero rotation is no longer on a symmetry axis. These locations are generally given by three curves in the two-dimensional plane. Nonzero rotation is found in the six zones divided by the zero curves and the sense of rotation is opposite in neighboring zones. One point of maximal rotation speed is found per zone, actual and maximal values of $\langle \omega \rangle$ will be discussed later.
To conclude our discussion of Motor 1, a few expressions for fixed values are given. For $\alpha = \pi/4$,
\begin{multline}
\oint  dl\,  r_{\shortparallel}^{2} = 
\frac{R}{36}  \biggl(\left(3+2 \sqrt{2}\right) R^2+6 (x+y) R
\\
+18 \left(\left(1+\sqrt{2}\right) x^2-2 y
   x+\left(1+\sqrt{2}\right) y^2\right)\biggr),
\end{multline}
\begin{multline}
\oint  dl\,  r_{\shortparallel}^{3} = 
\frac{R}{24}  (x-y) 
\\
\times \left( 6 (x+y) R + 6 \sqrt{2} \left(x^2+4 y x+y^2\right) - \sqrt{2} R^2\right),
\end{multline}
and for $\alpha = \pi/6$
\begin{multline}
\oint  dl\,  r_{\shortparallel}^{2} = 
\frac{R}{48}  \biggl(\left(5+2 \sqrt{3}\right) R^2+4 \left(3+\sqrt{3}\right) y R 
\\
+ 36 x^2+12 y \left(\left(3+2
   \sqrt{3}\right) y-2 \sqrt{3} x\right) \biggr),
\end{multline}
\begin{multline}
\oint  dl\,  r_{\shortparallel}^{3} = 
-\frac{R}{288 \sqrt{3}}  \biggl(\frac{\sqrt{3} R^3}{2}+18 y R^2+108 x y R 
\\
- 54 \sqrt{3} (x-y) (x+y) R-27 y \left(R^2+12 x^2-4 y^2\right)
\\
+27 \sqrt{3} x \left(R^2-4 x^2+12 y^2\right) \biggr).
\end{multline}

Similarly for Motor 2 we find, again choosing the origin of the coordinate system $(x, y)$ in the center of mass of the motor,
\begin{multline}
\oint  dl\,  r_{\shortparallel}^{2} = 
\frac{2 R}{9}  \biggl(R^2+3 R y \cos 3 \alpha +9 y^2 + 9 x^2 \sin \alpha  
\\
+\sin ^2 \alpha 
   \left(R^2 (3 \sin \alpha -2 \cos 2 \alpha )+ 9 x^2 -9 y^2\right) \biggr),
\end{multline}
while $I / M = R^2 (2 - \cos 2 \alpha) / 18 +x^2+y^2$.
Again
\begin{multline}
\oint  dl\,  r_{\shortparallel}^{3} = 
\frac{R}{3}  x \sin 2 \alpha
\biggl(R (3 y+R \cos \alpha ) (1 - 2 \cos 2 \alpha )
\\
+3 \left(x^2-3 y^2\right) \cos \alpha \biggr)  
\end{multline}
determines the sense of rotation and reflects the chiral symmetry of the motor. Specifically, putting the rotation axis on the symmetry axis of the motor, $x =0$, results in a zero average rotation. Putting the axis on symmetrical location across the symmetry axis yields the same rotation speed, but in opposite sense, $\langle \omega \rangle[-x,y] = - \langle \omega \rangle[x,y]$. Again, reducing the motor to a bar shaped object, $\alpha = 0$ or $\alpha = \pi/2$, a preferred sense of rotation can no longer be determined and $\langle \omega \rangle = 0$. Fig.÷ \ref{fig:contoursmotor2} shows  $\langle \omega \rangle$ for general positions of the rotation axis, for different configurations of the motor shape. For certain configurations the expressions simplify to reflect a higher symmetry. For $\alpha = \pi/6$ for example the shape is that of an equilateral triangle with side $2R$, with
\begin{equation}
\oint  dl\,  r_{\shortparallel}^{2} = \frac{1}{4} R \left(R^2+6 \left(x^2+y^2\right)\right),
\end{equation}
and
\begin{equation}
\oint  dl\,  r_{\shortparallel}^{3} =\frac{3}{4} R x \left(x^2-3 y^2\right),
\end{equation}
so that the full expression for the rotation speed becomes
\begin{equation}
\langle \omega \rangle
= v_{0} \frac{36 x \left(x^2-3 y^2\right)}{R^4+18 \left(x^2+y^2\right) R^2+72 \left(x^2+y^2\right)^2}.
\end{equation}
The rotation speed will be zero, when the rotation axis is on any of the three symmetry axes of the equilateral triangle, $x = \pm \sqrt{3} y$ besides $x = 0$, while the extrema are located on the vertices of a regular hexagon of side $\sqrt{(3 + \sqrt{33})/6} R$.
Finally, for Motor 2, the expressions  for $\alpha = \pi/4$ are
\begin{multline}
\oint  dl\,  r_{\shortparallel}^{2} = 
\frac{ R}{18} \biggl(\left(4+3 \sqrt{2}\right) R^2-6 \sqrt{2} y R
\\
+18 \left(\left(1+\sqrt{2}\right)
   x^2+y^2\right) \biggr),
\end{multline}
\begin{multline}
\oint  dl\,  r_{\shortparallel}^{3} = 
\frac{R x}{6} \left(\sqrt{2} R^2+6 y R+3 \sqrt{2} \left(x^2-3 y^2\right)\right).
\end{multline}

For Motor 3 the origin of the coordinate system $(x, y)$ is chosen in the center of the disk sector, and the expressions for a general shape ($\alpha$), size ($R$), and location $(x, y)$ with respect to the rotation axis are
\begin{multline}
\oint  dl\,  r_{\shortparallel}^{2} = 
\frac{R}{3}  \biggl(2 R^2+6 x^2 \sin ^2 \alpha +3 \left(x^2+y^2\right) \alpha 
\\
+ 6 y \cos \alpha  (y \cos
   \alpha -R)+3 (x-y) (x+y) \cos \alpha  \sin \alpha  \biggr),
\end{multline}
even for a $x \rightarrow -x$ transformation, and
\begin{multline}
\oint  dl\,  r_{\shortparallel}^{3} = 
\frac{2 R x}{3}  \biggl(\left(x^2-3 y^2\right) \sin 3 \alpha 
\\
- 3 R (R-3 y \cos \alpha ) \sin \alpha
    \biggr),
\end{multline}
odd for a $x \rightarrow -x$ transformation, so that the sense of rotation is inverted by putting the axis of rotation at its mirror location with respect to the symmetry axis, $x = 0$, of Motor 3. This is because the inertial moment,
\begin{equation}
I / M = \frac{R^2}{2}-\frac{4 R y \sin \alpha  }{3 \alpha }+x^2+y^2,
\end{equation}
is also even for the $x \rightarrow -x$ transformation.
For certain cases the expressions become simpler, such as for a semi-disk, $\alpha = \pi/2$:
\begin{equation}
\oint  dl\,  r_{\shortparallel}^{2} = 
\frac{R}{6}  \left(4 R^2+3 (4+\pi ) x^2+3 \pi  y^2\right),
\end{equation}
and
\begin{equation}
\oint  dl\,  r_{\shortparallel}^{3} =
-\frac{2  R x}{3} \left(3 R^2+x^2-3 y^2\right),
\end{equation}
and for $\alpha = \pi/3$:
\begin{multline}
\oint  dl\,  r_{\shortparallel}^{2} = 
\frac{R}{12}  \biggl(8 R^2-12 y R+\left(18+3 \sqrt{3}+4 \pi \right) x^2
\\
+\left(6-3 \sqrt{3}+4 \pi \right)
   y^2 \biggr),
\end{multline}
and
\begin{equation}
\oint  dl\,  r_{\shortparallel}^{3} =
\frac{\sqrt{3}}{2}  R^2 x (3 y-2 R).
\end{equation}
Fig.÷ \ref{fig:contoursmotor3} shows $\langle \omega \rangle$ for these and other configurations as a function of the position of the rotation axis. Again extrema are obtained in the $x,y$-plane at finite values, while three curves of zero rotation mark areas of opposite rotation.

\end{document}